\documentclass[12pt]{iopart}
\usepackage{epsfig}
\usepackage{graphicx}

\newcommand{\be}{\begin{equation}}
\newcommand{\ee}{\end{equation}}
\newcommand{\bd}{\begin{displaymath}}
\newcommand{\ed}{\end{displaymath}}

\begin{document}

\title[On the conditions for the existence of  Perfect Learning]{On the conditions for the existence of Perfect Learning and power law in learning from stochastic examples by Ising perceptrons}
\author{
T Uezu \footnote{Present address: 
 Department of Mathematics, UMIST, 
PO Box 88, Manchester M60 1QD.
e-mail: tatsuya@fire.ma.umist.ac.uk}
}
\address{  Graduate School of Human Culture, Nara Women's University, Nara 630-8506, Japan}
\begin{abstract}

In a previous letter, we studied
 learning from stochastic examples
by perceptrons with Ising weights in the framework of
statistical mechanics.  Under the one-step
replica symmetry breaking ansatz,
the behaviours of learning curves were
classified according to some
local property of the rules by which examples were drawn.
Further, the conditions for the existence
of the Perfect Learning together with other behaviors
of the learning curves were given.
In this paper, we give the detailed derivation about
these results and further argument about
the Perfect Learning together with extensive numerical calculations.

\end{abstract}


\newfont{\msamfnt}{msam10}
% \g \sim 
\newcommand{\gsim}{\mbox{$\;$ \msamfnt\symbol{'046} $\;$}}
% \l \sim 
\newcommand{\llsim}{\mbox{$\;$ \msamfnt\symbol{'056} $\;$}}

\def\veco{\mbox{\boldmath $w^o$}}
\def\vecx{\mbox{\boldmath $x$}}
\def\vecw{\mbox{\boldmath $w$}}
\def\veceta{\mbox{\boldmath $\eta$}}

\clearpage

\section{Introduction}
In the problem of learning from examples by feed forward
networks, learning curves of the generalization error 
$\epsilon _{g}$ have been calculated for various 
types of networks \cite{Rau 93}.
From these studies, it turned out that
when the number of examples $p$ is 
large relative to the number of synaptic weights $N$, 
that is,  when $\alpha=p/N$ is large,
the learning curves 
exhibit only a few types of behaviours \cite{Seung 92}-\cite{Amari 92}.
For example, learning curves of networks with continuous
weights all exhibit power laws 
\[ (\epsilon _{g} - \epsilon _{min}) \propto \alpha ^{- \gamma}, \]
where $\gamma$ depends on architectures, types of weight vectors
and so on.

On the other hand, 
in the learning behaviors for the case of discrete weights,
in addition to the power laws
it was shown that there exists the Perfect Learning(PL)
for deterministic and realizable cases \cite{Gyorgyi 90,Sompolinski 90}.
That is, learners' weight vectors coincide to the teacher's weight
vector at a finite $\alpha$.
Then, it is very interesting to clarify the existence conditions
for the PL in the case of discrete weights and under the
presense of exernal noise.\par
In the previous paper \cite{Uezu 96c}, we reported
about these conditions. The results are similar
to those by Seung \cite{Seung 95} who classified the learning behaviours
 of Ising networks by introducing two exponents $y$ and $z$. We gave the
other meaning of $y$ and $z$ and obtained
the relation $y=2z$.  Further, asymptotic
behaviours of learning curve were
also investigated.

The purpose of this paper is 
to give the detailed derivation on the conditions for the
existence of the Perfect Learning and 
 the asymptotic behaviours of 
learning curves in the problem
of learning from stochastic examples by perceptrons 
with Ising weights by using the replica method.\par
In the following section, we formulate the problem.
In \S3, we analyse the replica symmetric(RS) solution.
The conditions for the existence of the PL is given in \S4.
The one-step replica symmetry breaking(1RSB) solution
is studied in \S5.  The results of numerical calculations are given
in \S6.  \S7 is devoted to summary and discussion.

\section{Formulation}
We consider a stochastic target relation between $N$-dimensional
input vector $\vecx$ and binary output $r \in \{1,-1\}$
which is represented by a conditional probability $p_r(r| \vecx)$.
It is assumed that an input vector $\vecx$ is normalized as
$\mid \vecx \mid \ = \sqrt{N}$ and $p_r(r| \vecx)$ is a function
of the inner product between the input $\vecx$ and the optimal Ising
weight $\veco$ as
\begin{eqnarray}
  p_r(+1| \vecx) & = & {\cal P}(u^o)=\frac{1+P(u^o)}{2}, \\
  u^o & \equiv &
( \vecx \cdot \veco)/\sqrt{N}. \nonumber
\end{eqnarray}
We further assume that the function $ P(u)$ is not decreasing
w.r.t. $u$ and
behaves as
\begin{equation}
 P(u) \simeq a \; {\rm sgn}(u) |u|^{\delta}, \; \; (\delta \ge 0),
\end{equation}
near $u=0$.  Further, $P(-u)=-P(u)$ is assumed for brevity.\\
The case of $\delta=0$ corresponds to the output noise
model \cite{Opper 91} in which the  output of the target
perceptron is reversed to the opposite by noise with
a probability.   On the other hand, $\delta=1$ corresponds to
the input noise model \cite{Gyorgyi 90} in
which the input of the target perceptron is corrupted by
Gaussian noise with mean zero.\par
We assume that a set of $p$ examples 
$ \xi_p= \{ (\vecx ^1, r_1 ^o), (\vecx ^2, r_2^o),
 ..., (\vecx ^p, r_p^o) \} $
is obtained as follows.
$\vecx ^{\mu}$ is independently and uniformly drawn from a hyper sphere 
of radius $\sqrt{N}$ at the origin 
in the $N$-dimensional space and $r_{\mu} ^o$ is obtained with the conditional 
probability $p_r(r_{\mu} ^o|\vecx^{\mu})$ for each $\vecx^{\mu}$. 
For the given realization of examples $\xi_p$, 
the number of false predictions is given as 
\begin{equation}
E[ w ,\xi_p] = \sum_{\mu =1}^{p} \Theta (- r^o _{\mu}u_{\mu}), \;\;
 u_{\mu} \equiv ( \vecx^{\mu} \cdot \vecw)/\sqrt{N},
\label{eq:energy}
\end{equation}
where $\Theta(x)=1$ for $x \ge 0$ and $\Theta(x)=0$ for $x <0$.
The performance of the learning is evaluated by the generalization error 
$\epsilon_{g}$. 
This is expressed as
\begin{eqnarray}
\epsilon_{g} &=& <{\cal P}(u^o)(1-\Theta(u))+
(1-{\cal P}(u^o)) \Theta(u) > \\
 & = & \epsilon_{min}
+2 \int_{0}^{\infty} Dy P(y) H(\frac{Ry}{\sqrt{1-R^2}}),  \nonumber \\
\epsilon_{min} & = & \frac{1}{2}-\int_{0}^{\infty} Dy P(y), \nonumber
\end{eqnarray}
where $ < \cdots >$ represents the average over a novel example and 
$\epsilon_{min}$ is the minimum value of the generalization error 
obtained by the optimal weight $\vecw^o$.
$R$ is the overlap between the optimal weight vector and a
weight vector of a learner, $ R= (\veco \cdot \vecw)/N$.
Further, as usual, $Dy= {\rm exp}(-y^2 /2)dy/\sqrt{2\pi}$ and 
$H(x)=\int_x ^{\infty} Dy $.
  In particular, when $\Delta R \equiv 1-R$ is small, we obtain the relation
\begin{equation}
\Delta \epsilon _g \equiv (\epsilon _g - \epsilon_{min}) 
\simeq 
\frac{2s}{(1+\delta) \sqrt{2\pi}}
 {(2 \Delta R)^{\frac{1+\delta}{2}}},
\label{eq:smalldR}
\end{equation}
where $s  \equiv  a \int_0 ^{\infty} 
Dy y  ^{ 1+\delta }$.\par
In this paper, we adopt the Gibbs algorithm with temperature $T$
as a learning algorithm. 
The minimum-error algorithm,
which minimizes the number of false predictions 
on the presented examples,
 is obtained by taking $T \rightarrow +0$ limit.\par
From the energy defined by the equation (\ref{eq:energy})
the partition function $Z$ with the inverse temperature $\beta$ is 
given by 
\[ Z= {\rm Tr}_w  e ^{-\beta E[ w ,\xi_p]}
= {\rm Tr}_w \Pi_{\mu =1}^p 
[e ^{-\beta}+(1-e^{-\beta}) \Theta( r_{\mu}u_{\mu})],  \]
where Tr$_w$ implies the summation over all configurations of $\vecw$.
The average free energy $f$ per weight is calculated by
the standard recipe
\[ -\beta N f = < \ln Z > _{\xi_p, w^o}
= \lim_{n \rightarrow 0} \frac{1}{n}( < Z^n > _{\xi_p, w^o}-1),\]
where $< \cdots > _{\xi_p, w^o}$ denotes the average over
quenched variables.\par
$ < Z^n > _{\xi_p, w^o}$ becomes a function of several 
replica order parameters,
namely the overlap between weight vectors of learners
$q^{\alpha \beta}=\frac{(\vecw^{\alpha} \cdot \vecw^{\beta})}{N}$,
 its conjugate $\hat{q}^{\alpha \beta}$,
the overlap between the weight vector of a learner and the optimal weight
vector
$R^{\alpha}=\frac{(\veco \cdot \vecw^{\alpha})}{N}$ 
, and its conjugate $\hat{R}^{\alpha}$.  See Appendix A
for a derivation of the free energy.

\section{RS solution}
Let us cosider the RS solution.
For the RS solution, any quantity does not depend
on the replica indices and 
we put $q^{\alpha \beta}=q, \hat{q}^{\alpha \beta}=\hat{q}, 
R^{\alpha}=R$
and $\hat{R}^{\alpha}=\hat{R}$.  Then, the RS free energy $f_{RS}$ becomes
\begin{eqnarray}
\hspace*{-10mm}
&&  -{\beta} f_{RS}(q,\hat{q},R,\hat{R},\beta) =
  -\frac{\hat{q}}{2} (1- q) - R \hat{R} + \alpha K + I,\\
\hspace*{-10mm}
&& \hspace{0.5cm} K \equiv \int Dy2{\cal P}(y) \int Du \ln
 \tilde{H}(\frac{\sqrt{q-R^2}u-Ry}{\sqrt{1-q}})  
  = \int Du \ln \tilde{H}(u/Q) E(u/Q),  \\
\hspace*{-10mm}
&& \hspace{0.5cm} I \equiv \int Dt \ln [2 {\rm cosh}( \sqrt{\hat{q}}t+ \hat{R})],  \\
\hspace*{-10mm}
&&\hspace{0.5cm}  E(u)= \int Dy 2 {\cal P}(- \Lambda) 
 =1-e^{-v^2/2}\int _0 ^{\infty}Dy(e^{-vy}- e^{vy})P(\xi y), \\
\hspace*{-10mm}
&& \hspace{0.5cm} \tilde{H}(u) \equiv e^{-\beta}+(1-e^{-\beta})H(u),
\;\; \Lambda = \xi y + \sqrt{1-\xi^2}Qu = \xi(y-v),\nonumber \\
\hspace*{-10mm}
&& \hspace{0.5cm} \xi=\sqrt{1-\frac{R^2}{q}},\;\;
Q=\sqrt{\frac{1-q}{q}},  v= -\frac{R}{\sqrt{q} \chi} u,
\;\; \chi =\frac{\xi}{Q}. \nonumber
\end{eqnarray}

\subsection{ Saddle point equations(S.P.E.)}

The saddle point equations are given by
\begin{eqnarray}
\hspace*{-10mm}
q & = & \int Du \tanh ^2 ( \sqrt{\hat{q}} u +\hat{R}),\label{eq:s1}\\
\hspace*{-10mm}
R & = & \int Du \tanh  ( \sqrt{\hat{q}} u +\hat{R}),\label{eq:s2}\\
\hspace*{-10mm}
\hat{q} & = & \frac{\alpha Q}{1-q}
\int \tilde{D}u ( \tilde{\varphi}(u)) ^2 E(u),\label{eq:s3}\\
\hspace*{-10mm}
\hat{R} & =  & -  \frac{\alpha}{\sqrt{q-R^2}} 
\int \tilde{D}u \tilde{\varphi}(u)
 \int Dy y 2 {\cal P}(\Lambda) 
 =  -  \frac{\alpha}{\sqrt{q-R^2}}
\int \tilde{D}u \tilde{\varphi}(u)  w(u), \label{eq:s4}\\
\hspace*{-12mm}
&&\hspace{0.5cm} w(u)\equiv \int Dy y P(\Lambda)
 =  e^{-v^2/2}  \int_0 ^{\infty} 
 Dy P(\xi y) [ (y+v) e^{-vy}+ (y-v)e^{vy}],\label{eq:s5} \\
\hspace*{-10mm}
&&\hspace{0.5cm}  \tilde{D}u = 
\frac{du}{\sqrt{2 \pi}} e^{- Q^2 u^2 /2},\;\;
\tilde{\varphi}(u)= \frac{\tilde{H}'(u)}{\tilde{H}(u)}.
 \nonumber
\end{eqnarray}
For later use, we give the expression 
of the entropy $S_{RS}$,
\begin{eqnarray}
S_{RS} & = & -\beta f_{RS} - \alpha \beta e^{-\beta} J, \\
&& \hspace{0.5cm} J = \int Dy2{\cal P}(y) \int Du 
\frac{H(\frac{\sqrt{q-R^2}u-Ry}{\sqrt{1-q}})-1}
{\tilde{H}(\frac{\sqrt{q-R^2}u-Ry}{\sqrt{1-q}})}. \nonumber 
\end{eqnarray}
Defining $L$ as $L = K -  \beta e ^{-\beta} J$,
$S_{RS}$ becomes
\begin{equation}
S_{RS}= - \frac{\hat{q}}{2}(1-q) -R \hat{R}+ I  + \alpha L,
\end{equation}
where $L$ is expressed as
\begin{equation}
L= \int Du E(u/Q) \{ \ln [ 1+ (e ^{\beta} -1) H(u/Q) ] -\beta 
\frac{H(u/Q)}{\tilde{H}(u/Q)} \}.
\end{equation}
Also, the energy(training error) per weight is  expressed as
\begin{equation}
e_t = -\alpha e^{-\beta} J.
\end{equation}

\subsection{Numerecal calculations of S.P.E. for the RS solution}
Here, we show the results of numerical calculations for the RS solution.

\noindent
(I) $\delta =1$ \par
We treated $P(y)=1-2H(y)$, in which $\epsilon_{min}=\frac{1}{4}$.
 In Figure 1, for $T=1$
we depict $\alpha$ dependence of $q$, $R$, $S_{RS}$, and 
$\Lambda _1$ and $\Lambda _3$ which are indicators of AT-stability.
The RS solution is stable only when
 both $\Lambda _1$ and $\Lambda _3$ are negative.
From the numerical results, it seems
that as $\alpha \rightarrow \infty$, $q$ and $R$ tend to 1.
  In this case, the entropy $S_{RS}$ becoms zero at 
some value of $\alpha$, $\alpha_s(T)$,
and $\Lambda _3$ becomes zero at another value of $\alpha$,
 $\alpha_{AT}(T)$, for any $T$.  

\noindent
(II) $\delta =0$ \par
For the numerical calculations, 
we treated $P(y)=\frac{1}{2} $ sgn $(y)$, 
in which $\epsilon_{min}=\frac{1}{4}$.
In Figures 2-4,  for several temperatures 
we depict $\alpha$ dependence of $q$, $R$, $S_{RS}$, 
$\Lambda _1$ and $\Lambda _3$.
The most interesting feature is that there
is no solutions in which $q$ and $R$ tend to 1 
as $\alpha \rightarrow \infty$.
There are two branches of solutions.
We call them the branch I and the branch II.
Each solution is characterized by the behaviour in the limit of $\alpha 
\rightarrow 0$.  In the branch I, $q$ and $R$ tend to 0. 
On the other hand, in the branch II,  $q$ and $R$ tend to 1. 
 We attach the superscript I or II to any  quantity 
 estimated in the branch I or II, respectively.
From these figures, we note that when $T$ is greater than
 some  temperature,  say $T_s$,
 solutions in the both branches are AT-stable and their entropies are positive.
When $T<T_s$, the entropy of the branch II, $S^{II}_{RS}$,
becomes negative for any $\alpha$.  
Thus, $T_s$ is determined by the condition that
$S^{II}_{RS}$ changes its sign at small value of $\alpha$.
There exists a critical value of $\alpha = \alpha _s(T)$
at which the entropy of the branch I, $S^{I}_{RS}$, becomes 0.
Then, $S^{I}_{RS} >0 $ for $\alpha < \alpha _s(T)$
and $S^{I}_{RS} < 0 $ for $\alpha > \alpha _s(T)$.
Also, we note that for the branch I, the AT-instability
takes place at $\alpha =\alpha _{AT}(T)$ when
$T$ is smaller than some temperature, say, $T_{AT}(<T_s)$.
For $T< T_{AT}$, 
$\Lambda _3 ^{I} $  is positive for $\alpha > \alpha _{AT}(T)$,
whereas $\Lambda _3 ^{II} $ is positive for any $\alpha$.
On the other hand, $\Lambda _1 ^I $ and $\Lambda _1 ^{II} $
are always negative for any $T$ and $\alpha$.

\noindent
(III) general $\delta$ \par
The numerical calculations were performed for several
values of $\delta$ and $T$.  For example,
when $T=5$ and $\delta =0.3$, 
$q$ and $R$ tend to 1 as $\alpha \rightarrow 0$. See Figure 5(a). 
In  this case, the RS solution is AT stable and its
 entropy is positive.  There exists the other case in which
$q$ and $R$ tend to 1 as $\alpha \rightarrow \infty$.
See Figure 5(b).
In this case, $S_{RS}$ decreases and becomes 0 at 
finite $\alpha$, $\alpha_s(T)$,
for any $T$. $\Lambda _3 $ also becomes 0 at finite $\alpha$,
$\alpha_{AT}(T)$.\\

Within our numerical calculations, for any value of
$\delta$ and for $T<T_{AT}$, we obtain the relation
$\alpha _{AT}(T) > \alpha_{s}(T)$, and
 $\alpha _{AT}(T)$ and $ \alpha_{s}(T)$ are increasing as $T$ increases,
as long as these quantities are defined.
Further, except for the case of $\delta=0$, 
$\alpha _{AT}(T)$ and $ \alpha_{s}(T)$ increase as $\delta $ 
increases.\par
For any value of $\delta$, the entropy becomes negative
for small $T$.  Thus, in this case we have to consider 
the  replica symmetry breaking ansatz.

\subsection{Asymptotic relations for $\hat{q}$
and $\hat{R}$ when  $q \rightarrow 1$ and $R \rightarrow 1$ }
In this section, in order to derive the asymptotic
learning curves and to discuss the conditions 
for the existence of the PL, we summarize asymptotic relations for
$\hat{q}$ and  $\hat{R}$
when $q \rightarrow 1$ and $R \rightarrow 1$ 
for any values of $\alpha,
\beta$ and $\chi$ by evaluating equations ($\ref{eq:s3}$) and ($\ref{eq:s4}$)
under these limits.  
See  Appendix B for the derivation.\\
 $\hat{q}$ and $\hat{R}$ are estimated as follows.
For $0 < \beta < \infty$
\begin{eqnarray}
\hat{q} & \simeq & \frac{\alpha}{ \sqrt{\Delta q}} 
g_{1, \delta}(\chi , \beta) \mbox{ for $\delta \ge 0$},
\label{eq:a}\\
\hat{R} & \simeq & \alpha \frac{\xi^{ \delta}} {\sqrt{\Delta q}}
 g_{2, \delta}(\chi , \beta) \mbox{ for $\delta \ge 0$}. \label{eq:b}
\end{eqnarray}
For $\beta =\infty$,
\begin{eqnarray}
\hat{q} & \simeq & \frac{\alpha}{ (\Delta q)^2}
g_{3}\;\;
 \hbox{ for $\delta >0$, or for $\delta=0$ and $k<1$},\\
 &  \simeq & \frac{\alpha}{ \sqrt{\Delta q}}
g_{3,D} \mbox{ for deterministic case}, \\
% case of \hat{R}
\hat{R} & \simeq &  \frac{\alpha}{\Delta q}
g_{4} \mbox{ when  $P(y)$ is not constant for $y>0$},\\
&  \simeq &  \frac{\alpha}{\Delta q}
 \frac{2k \xi^3}{\pi} \mbox{ when $P(y)\equiv k$ for $y>0$ and $k<1$},\\
 &=\hat{q}&   \simeq \frac{\alpha}{ \sqrt{\Delta q}}
g_{3, D}  \hbox{ for deterministic case}.
\end{eqnarray}
Expressions for $g$s are given as follows.
\begin{eqnarray*}
g_{1, \delta}(\chi , \beta) & \equiv &
 \frac{1}{ \sqrt{2\pi}} \int du  \tilde{\varphi} (u) ^2
 E_{\delta}(u, \chi) ,\\
&& \hspace{0.5cm} E_{\delta}(u, \chi)
=1 \;\;{\rm for }\; \delta >0, \;\;
 E_0(u, \chi)=1-k+2k H(u/\chi), \\
 g_{2, \delta}(\chi , \beta) & \equiv &
 \frac{a \delta }{\sqrt{2\pi}}
 (1 - e^{-\beta}) \frac{1}{\chi} (1+ \chi^{-2})^{(\delta -1)/2} \\
&& \times \int_0 ^{\infty}Dz z ^{\delta -1} 
\int_{-\infty} ^{\infty} Dt 
[ \frac{1}{\tilde{H} (\frac{\chi t -z}{\sqrt{1 + \chi ^2}})}
 + \frac{1}{\tilde{H} (-\frac{\chi t -z }{\sqrt{1 + \chi ^2}})} ],
 \mbox{ for $\delta >0$}, \\
 g_{2,0}(\chi, \beta) & = &
 \frac{k(1 - e^{-\beta})}{\pi }
 \frac{1}{\sqrt{1+ \chi^2}}
 \int_{-\infty} ^{\infty} Dx
\frac{1}{\tilde{ H}(\frac{\chi x}{\sqrt{1+\chi ^2}})},\\
 g_{3} &  \equiv &
\int_0 ^{\infty} Dy  y^2 [ 1 - P(y) ], \;\;
g_{3,D}  \equiv  \frac{2}{\sqrt 2 \pi}\int Du \frac{h(u)}{H(u)},\\
g_4 & \equiv & \int _0 ^{\infty} Dy P(y)(y^2-1).
\end{eqnarray*}
Now, let us see the behaviours of $g$s for later use.
$g_{1, \delta}(\chi , \beta)$ is finite for $0<\beta<\infty$ and for 
$\delta \ge 0$ for any $\chi$, since $H(x)$ is bounded.
As for $g_{2, \delta}(\chi , \beta)$, we obtain for $0<\beta< \infty$
and for $\delta>0$,
 \begin{eqnarray}
g_{2,\delta}(\chi, \beta)
 & \sim & \nu _0 \chi ^{-\delta}, \mbox { for $\chi \ll 1$}, \label{eq:g2d1}\\
& \sim & \nu _1 \chi ^{-1}, \mbox { for $\chi \gg 1$},\label{eq:g2d2}\\
& \sim & \mbox { finite for finite $\chi$},\label{eq:g2d3}
\end{eqnarray}
where $\nu_0 =  \frac{a \delta (1 -e^{-\beta})} {\sqrt{2\pi}}
\int_0 ^{\infty}Dz z ^{\delta -1}
\{ \frac{1}{\tilde{H}(z)} + \frac{1}{\tilde{H}(-z)}\}$ and 
$\nu_1 =  \frac{2 s \beta} {\sqrt{2\pi}}.$  In the case of $\delta=0$, 
for $0<\beta< \infty$,
  $g_{2, 0}$ is finite except for $\chi \gg 1$, that is,
 \begin{eqnarray}
 g_{2, 0} & \sim &  \frac{2k }{\pi} \tanh (\beta /2)
\mbox { for $\chi \ll 1$},\\
& \sim &  \frac{k \beta }{\pi} \chi^{-1} \mbox { for $\chi \gg 1$},\\
& \sim & \mbox { finite for finite $\chi$}.
\end{eqnarray}
On the other hand, quantities for $\beta = \infty$, $g_{3}, g_{3, D}$
and $g_4$ are all finite.\par
Here, for later use, we give the 
asymptotic form of the entropy $S_{RS}$.
 $L$ is evaluated as
\begin{eqnarray}
L & = & \sqrt{\Delta q}\; r(\chi, \beta), \\
r(\chi, \beta) & \equiv & \frac{1}{\sqrt{2 \pi}}
\int du E_{\delta}(u, \chi)
 \{ \ln [ 1 + (e ^{ \beta} -1) H(u) ]
- \beta \frac{H(u)}{ \tilde{H}(u)} \}.
\end{eqnarray}
$r(\chi, \beta) $ is finite for $0<\beta < \infty$ and for any $\delta$ and 
any $\chi$.
Then, as $ q \rightarrow 1$ and  $ R \rightarrow 1$, for $0< \beta < \infty$
and for any $\delta$ and any $\chi$, 
$S_{RS}$ is expressed as follows.
\begin{equation}
S_{RS} =-\frac{\hat{q} \Delta q}{2}
 -(1 - \Delta R) \hat{R}
+ \alpha \sqrt{\Delta q}\; r(\chi,\beta) + I.
\end{equation}

\subsection{Asymptotic solutions of S.P.E. when 
$q \rightarrow 1$ and $R \rightarrow 1$.}
The equations for $\Delta q = 1 -  q$ and $ R$ are obtained by
\[
 \Delta q = 2 \frac{\partial I}{\partial \hat{q}},
\; \;  R =  \frac{\partial I}{\partial \hat{R}}.
\]
Thus, we have to estimate $I$ in the  asymptotic region.
$I$ behaves differently according to the values of 
$\mu\equiv \frac{\hat{R}}{2 \hat{q}}$. 
We give the expressions for $I$ in Appendix C. \par

After several algebra, we obtain the following asymptotic behaviours.
\begin{enumerate}
\item  $ \delta >\frac{1}{3}$.\\
For $ \alpha \gg 1$,
\begin{eqnarray*}
\mu & \simeq & \mu_0 (\frac{\ln \alpha}{\alpha})^{\frac{2\delta}{3\delta -1}},
\;\; \mu _0 = [ \frac{\delta}{\hat{q}_0(3\delta -1)} ]
^{\frac{2\delta}{3\delta -1}}\\
\Delta q & \simeq & q_0 \mu_0^{\frac{1+\delta}{\delta}}
 (\frac{\ln \alpha}{\alpha})^{\frac{2(1+\delta)}{3\delta -1}},\;\;
\Delta R  \simeq  R_0 \mu_0^{1/\delta}
 (\frac{\ln \alpha}{\alpha})^{\frac{2}{3\delta -1}},\\
\hat{q} & \simeq & \hat{q}_0 \mu _0^{-\frac{1+\delta}{2 \delta}}
\alpha ^{\frac{4\delta}{3\delta -1}}
 (\ln \alpha)^{-\frac{1+\delta}{3\delta -1}},\;\;
\hat{R}  \simeq  \hat{R}_0 \mu _0^{-\frac{1-\delta}{2 \delta}}
\alpha ^{\frac{2\delta}{3\delta -1}}
(\ln \alpha)^{-\frac{1-\delta}{3\delta -1}}.
\end{eqnarray*}
\item $ \delta =\frac{1}{3}$.\\
For $ \alpha \gg 1$,
\begin{eqnarray*}
\mu & \simeq & \mu_0 \alpha^{-\frac{1}{6}}
  e^{-\frac{2}{3} \hat{q}_0 \alpha},
\;\; \mu _0 = [ \frac{1}{2 \sqrt{\hat{q}_0}} ]^{\frac{1}{3}},\\
\Delta q & \simeq & q_0 \mu_0^4
 \alpha^{-\frac{2}{3}} e^{-\frac{8}{3} \hat{q}_0 \alpha},\;\;
\Delta R  \simeq  R_0 \mu_0^3  
 \alpha^{-\frac{1}{2}}e^{-2 \hat{q}_0 \alpha},\\
\hat{q} & \simeq & \hat{q}_0 \mu _0^{-2}
 \alpha^{\frac{1}{3}}e^{\frac{4}{3} \hat{q}_0 \alpha},\;\;
\hat{R}  \simeq  \hat{R}_0 \mu _0^{-1}
 \alpha^{\frac{1}{6}}e^{\frac{2}{3} \hat{q}_0 \alpha}.
\end{eqnarray*}
\item $ 0< \delta <\frac{1}{3}$.\\
For $ \alpha \ll 1$,
\begin{eqnarray*}
\mu & \simeq & \mu_0 (\frac{\ln \frac{1}{\alpha}}{\alpha})
^{-\frac{2\delta}{1-3\delta }},
\;\; \mu _0 = [ \frac{\delta}{\hat{q}_0(1-3\delta )} ]
^{- \frac{2\delta}{1-3\delta}}\\
\Delta q & \simeq & q_0 \mu_0^{\frac{1+\delta}{\delta}}
 (\frac{\ln  \frac{1}{\alpha}}
{\alpha})^{-\frac{2(1+\delta)}{1-3\delta}},\;\;
\Delta R  \simeq  R_0 \mu_0^{1/\delta}
 (\frac{\ln  \frac{1}{\alpha}}{\alpha})^{-\frac{2}{1-3\delta}},\\
\hat{q} & \simeq & \hat{q}_0 \mu _0^{-\frac{1+\delta}{2 \delta}}
\alpha ^{- \frac{4\delta}{1-3\delta }}
 (\ln  \frac{1}{\alpha})^{\frac{1+\delta}{1-3\delta }},\;\;
\hat{R}  \simeq  \hat{R}_0 \mu _0^{-\frac{1-\delta}{2 \delta}}
\alpha ^{-\frac{2\delta}{1-3\delta}}
 (\ln \frac{1}{\alpha})^{\frac{1-\delta}{1-3\delta}}. 
\end{eqnarray*}
\item $ \delta =0$.\par
(a) $0 \llsim \mu <1$.

For $\alpha \ll 1$,
\begin{eqnarray*}
\Delta R & \simeq & \frac{1}{ \sqrt{2 \pi}} \psi(\mu)
\alpha ^2 ( \ln \frac{1}{ \alpha} )^{-2}, \; \; \Delta q=2\mu \Delta R, \\
\hat{R} & \simeq & \frac{2}{ \mu}
\ln [ \frac{1}{\alpha } ( \ln \frac{1}{\alpha })^{3/4} ],
\; \; \hat{q}=\frac{\hat{R}}{2\mu}.
\end{eqnarray*}
$\chi$ and $\mu$ are determined by the following equations.
\begin{eqnarray}
\frac{g_{2, 0}(\chi, \beta)}{2 g_{1,0}(\chi)}
& =& \frac{1}{1 + \chi ^2},\label{eq:rschia}\\
\mu& =&  \frac{1}{1+\chi ^2}\label{eq:rsmua}.
\end{eqnarray}

(b) $1\le \mu $.

For $\alpha \ll 1$,
\begin{eqnarray*}
\Delta R & \simeq & 2 a_{\mu} 
( \frac{1}{\alpha} \ln \frac{1}{ \alpha} )^{-2},
 \; \; \Delta q=2 \Delta R, \\
\hat{R} & \simeq & \frac{2 \mu }{2 \mu -1}
\ln ( \frac{1}{\alpha } \ln \frac{1}{\alpha } ),
\; \; \hat{q}=\frac{\hat{R}}{2\mu}.
\end{eqnarray*}
$\mu$ is determined by
\begin{equation}
\mu =  \frac{g_{2,0}(\chi=0, \beta)}{2g_{1,0}(\beta)}.\label{eq:rsmub}
\end{equation}
\end{enumerate}

\vspace{8mm}Now, let us check the validity of the above solutions.
First, let us see the case of $\delta > 0$.  We only have to
see the conditions for $\mu \ll 1, \Delta R \ll 1$ and 
$\hat{R} \gg 1$. For $\delta >1/3$, the conditions are 
\begin{eqnarray*}
&& g_{1,\delta} ^{(\delta-1)/\delta}\beta^{1/\delta} \gg 
 \frac{\ln \alpha}{ \alpha }
\;\;\; (\mu \ll 1),\\
&&g_{1,\delta} ^{-2}\beta^3 \gg  \frac{\ln \alpha}{ \alpha }
\; \;\; (\Delta R \ll 1),\\
&&g_{1,\delta} ^{2(\delta-1)}\beta^2 \gg 
\frac{(\ln \alpha)^{1 - \delta}}{\alpha^{2\delta}} \;\;\; (\hat{R} \gg 1).
\end{eqnarray*}
These are satisfied if $\alpha \gg 1$.
For $\delta =1/3$, the condition is $\hat{q}_0>0$
and automatically satisfied for $\beta >0$.
 For $0<\delta <1/3$, the conditions are
\begin{eqnarray*}
&& g_{1,\delta} ^{(\delta-1)/\delta}\beta^{1/\delta} \ll 
\frac{\ln \frac{1}{\alpha}}{ \alpha }
\;\;\; (\mu \ll 1),\\
&&g_{1,\delta} ^{-2}\beta^3 \ll
\frac{\ln \frac{1}{\alpha}}{ \alpha }  \;\;\; (\Delta R \ll 1),\\
&&g_{1,\delta} ^{2(\delta-1)}\beta^2 
\ll \frac{(\ln \frac{1}{\alpha})^{1 - \delta}}{\alpha^{2\delta}} 
\;\;\; (\hat{R} \gg 1).
\end{eqnarray*}
These are satisfied if $\alpha \ll 1$.
Therefore, no extra condition is necessary for the case of
 $\delta > 0$.  On the other hand, for $\delta =0$,
in the case of (a) the condition is 
that there is a positive solution $\chi$ of
 equation $(\ref{eq:rschia})$, and in the case of 
(b) the condition is $\mu \ge 1$, where $\mu$ is defined 
by equation $(\ref{eq:rsmub})$.
When $\beta \ll 1$, $\mu$ is estimated as
\begin{eqnarray*}
\mu & \sim & \frac{k \sqrt{2}}{\beta \sqrt{1 + \xi ^2}}
 \; \mbox{ for the case (a)}\\
 & \sim & \frac{k \sqrt{2}}{\beta} \; \mbox{ for the case (b)}.
\end{eqnarray*}
Thus, the case (a) is impossible for high temperatures.  On the other hand,
when $\beta \gg 1$, since $g_{1,0}$ becomes very large 
and  $g_{2,0}$ remains bounded in the both cases,
we obtain $\mu \ll 1 $ . Therefore, the case (b) is impossible
for low temperatures.\vspace{5mm}\\
The results obtained in this section suggest 
that for $\delta <1/3$
the PL exists and the solution with $q<1$ exists 
only in the finite region of $\alpha$, and for  $\delta \ge 1/3$ 
the PL does not exist and the solution with $q<1$ exists
 for any $\alpha$.  However, as is shown in the next two sections,
this is not correct. One reason is that the entropy of the RS solution becomes
negative for $T \rightarrow 0$ when $\alpha$ can be large enough. 
The other reason is that the condition $\delta < 1/3$ is 
different from the existence condition for the PL.\par
In the next section, we investigate 
the necessary and sufficient conditions for the existence
of the Perfect Learning.  
\section{Perfect Learning}
In the Perfect Learning, the weight vector of a student
coincides with the optimal weight vector at a finite value of $\alpha$,
$\vecw = \veco$.  In this case, $q=1$ and $R=1$.  
From equation ($\ref{eq:s1}$) and ($\ref{eq:s2}$)
the necessary and sufficient 
conditions for $q=1$ and $R=1$ at finite value of 
$\alpha$ are
\begin{equation}
\hat{R} \rightarrow \infty \mbox{ and }
\tau \equiv \frac{\hat{R}}{\sqrt{\hat{q}}}\rightarrow \infty.
\end{equation}
In the case of the PL, we impose the further 
condition $q=R$ 
when the limits $q \rightarrow 1$ and  
$R \rightarrow 1$ are taken,
because in the Perfect Learning, the teacher
and a student coincide.
Therefore, we have $\chi =\sqrt{\frac{q-R^2}{1-q}}= 
\sqrt{q} =1 $ and then $\xi = Q = \sqrt{\Delta q}$.  Thus, 
for $0< \beta <\infty$ we obtain
from equations ($\ref{eq:a}$), and ($\ref{eq:b}$),
\begin{eqnarray}
\hat{q} & = & \frac{\alpha}{\sqrt{\Delta q}} 
g_{1, \delta}(\chi = 1, \beta),\label{eq:4a}\\
\hat{R} & = & \alpha (\Delta q) ^{ (\delta-1)/2} 
g_{2, \delta}(\chi = 1, \beta) \label{eq:4b}.
\end{eqnarray}
For $ \beta =\infty$,
\begin{eqnarray}
\hat{q} & \simeq & \frac{\alpha}{ (\Delta q)^2}
g_{3}\; \mbox{ for $\delta >0$ or $\delta=0$ and $k<1$},\label{eq:4c}\\
 & \simeq &  \frac{\alpha}{ \sqrt{\Delta q}} g_{3,D}
 \hbox{ for deterministic case},\label{eq:4d}\\
\hat{R} & \simeq & \frac{\alpha}{\Delta q}
g_{4} \mbox{ when  $P(y)$ is not constant for $y>0$},\label{eq:4e}\\
&  \simeq &  \frac{\alpha}{\Delta q}
\frac{2k \xi^3}{\pi} \propto \alpha \sqrt{\Delta q}
\mbox{ when $P(y)\equiv k$ for $y>0$ and $k<1$},\label{eq:4f}\\
 &=\hat{q}&   \simeq \frac{\alpha}{ \sqrt{\Delta q}}
g_{3, D}  \hbox{ for deterministic case}.\label{eq:4g}
\end{eqnarray}
Let us see what are derived from these conditions.
First, let us consider the case of $0 < \beta<\infty$. 
In this case, both of $g_{1, \delta}(\chi=1, \beta)$ and 
$g_{2, \delta}(\chi=1, \beta)$ are finite for $\delta \ge0$. 
See subsection 3.3. Thus, 
from (\ref{eq:4a}) and (\ref{eq:4b}), the conditions for
$\hat{q}\rightarrow \infty, \hat{R}\rightarrow \infty$
 and $\tau \rightarrow \infty$
as $ q \rightarrow 1$ for any $\alpha$ 
are derived as
\begin{eqnarray}
\hat{q}& \rightarrow & \infty 
\mbox{ for any $\delta \ge 0$ and $0<\beta <\infty$},\\
\hat{R}&  \rightarrow & \infty  
\mbox{ for $0\le \delta<1$ and $0<\beta< \infty$},\\
\tau & \simeq & 
\alpha ^{1/2} (\Delta q) ^{(2\delta -1)/4}
\frac{g_{2, \delta}}{\sqrt{g_{1, \delta}}} \nonumber \\
&& \rightarrow \infty\mbox{ for $0 \le 
\delta < 1/2$ and $0<\beta < \infty $}.
\end{eqnarray}
Hence, the condition for the PL is $0 \le \delta < 1/2$. 
Next, let us consider the case of $\beta=\infty$.
When $P(y)$ is not constant for $y>0$, $g$s are finite. 
Then, from (\ref{eq:4c}) and (\ref{eq:4e}),
we obtain that both $\hat{q}$ and $\hat{R}$
tend to infinity and 
 $\tau$ becomes  $\tau \simeq \sqrt{\frac{\alpha}{g_3}} g_4$ and is finite.
Thus, in this case the PL does not exist.  On the other hand, 
when $P(y)\equiv k <1$ for $y>0$, 
$\hat{R} \propto \alpha \sqrt{\Delta q}$ tends to 0.
  Thus, the PL does not exist.
Finally, in the deterministic case, from (\ref{eq:4d})
and (\ref{eq:4g}), $\hat{q}=\hat{R}$ and $\tau =\sqrt{\hat{q}}$ 
tend to infinity since $g_{3,D}$ is finite. Hence, the PL exists.\par
Therefore, summarizing the above results,
we conclude that the PL exists 
for $ 0 \le \delta <1/2$ and $0< \beta < \infty $, and for  
the deterministic case.

As for the entropy $S_{PL}$ and the free energy $f_{PL}$
for the PL, we obtain the following reasonable results,
\[ S_{PL}=0,\;\; f_{PL}=\alpha \epsilon_{min}. \]
See Appendix D. 

\section{1RSB solution}
Although we adopt the Gibbs algorithm as the learning
strategy, we also have interest in the minimum-error
algorithm.  In the minimum-error algorithm 
we have to choose weights with minimum errors,
and on that account we only have to
take the limit $T \rightarrow +0$.
However, as is shown in \S 3
by numerical calculations, in the RS solution,
the entropy becomes negative for small $T$.
Thus, we have to
consider the breaking of the replica 
symmetry \cite{Parisi 80}.
In the one-step RSB solution, 
the matrix $q^{\alpha\beta}$ is divided into
$(n/m)^2$ small matrices with the dimension $m \times m$. 
The components of each off-diagonal matrix
are all $q_0$ and the components of 
each diagonal matrix are $q_1$ except for
 diagonal components with the value 0.
Likewise, $\hat{q}_{0}$ and $\hat{q}_{1}$ 
are defined for the matrix $\hat{q}^{\alpha\beta}$.
Further, $R^{\alpha}=R$ and 
$\hat{R}^{\alpha}=\hat{R}$ are assumed.
Then, the one-step RSB free energy $f_{1RSB}$ is derived to be
\begin{eqnarray}
\lefteqn{
-{\beta} f_{1RSB}(q_0,\hat{q_0},q_1,\hat{q_1},
R,\hat{R},m,\beta) }
\nonumber \\
& =& -\frac{\hat{q_1}}{2} (1- q_1)
+\frac{m}{2}(\hat{q_0}q_0-\hat{q_1}q_1) - R \hat{R} \\
&& + \frac{\alpha}{m}\int Dy2{\cal P}(y) \int Dz_0 \ln
\int Dz_1 [ \tilde{H}(\frac{\sqrt{q_0-R^2}z_0
+\sqrt{q_1-q_0}z_1-Ry}{\sqrt{1-q_1}}) ]^m \nonumber \\ 
&&  + \frac{1}{m} \int Dz_0 \ln \int Dz_1 [2 {\rm cosh}
( \sqrt{\hat{q_0}}z_0
+\sqrt{\hat{q_1} -\hat{q_0}}z_1+\hat{R})
 ]^m. \nonumber 
\end{eqnarray}
Further, according to Krauth-M\'{e}zard \cite{Krauth 89},
 we take the limits
$q_1 \rightarrow 1$ and $\hat{q_1} \rightarrow
\infty$.  Then, we obtain
\begin{equation}
f_{1RSB}(q_0,\hat{q_0},q_1=1,\hat{q_1}=\infty,R,
\hat{R},m,\beta)
= f_{RS}(q_0,m^2 \hat{q_0},R,m\hat{R},\beta m).
\end{equation}
From this relation, the equations for 
$q_0,\hat{q_0},R,\hat{R}$ and $m$ become the coupled 
equations of
the saddle point equations for the RS solution and 
the equation of $S_{RS}=0$, where 
$S_{RS}$ is the entropy for the RS 
solution.
Let us denote the solutions of these coupled equations by
$q=q_{c}, \hat{q}=\hat{q}_{c}, 
R=R_{c}, \hat{R}= \hat{R}_c$ and $ \beta=\beta_{c}$.
Then, the one-step RSB solutions are expressed by
$q_{0}  =  q_{c}, \hat{q}_{0}  = 
 (\frac{\beta}{\beta_{c}})^2 \hat{q}_{c} ,\;
\;  R  =  R_{c} ,\; 
\hat{R} =  \frac{\beta}{\beta_{c}} \hat{R}_c$ and
$m  =  \frac{\beta_{c}}{\beta}$.
Thus, to obtain the $ T \rightarrow +0$ limit we only
have to know the solution at $T=T_c \equiv \beta_c ^{-1}$.

\subsection{Numerical calculation of S.P.E. for the 1RSB solution }

\noindent
(I) $\delta>0$

As a special case, we treated $P(y)= 1 - 2H(y)$,
that is the case of $\delta=1$.
This is the same function as that calulated for the RS solution.
In Figure 6, we show $\alpha$ dependence of $T_c$.
See Figure 11 for $\alpha$ dependence of $ q_0, R$ and $\Delta \epsilon _g$.
As is seen from these figures, 1RSB solution seems to extend to
$\alpha=\infty$.
To study asymptotic behaviors, 
assuming the following relations for several quantities,
we estimated the coefficients $a_i$ and
exponents $b_i$ by the least square methods.
That is, for $\Delta q$, $\Delta R$,
$\Delta \epsilon _g$ and $T_c$ we assume
\begin{eqnarray*}
\ln A & = &  a_1 +b_1 \ln \alpha,\\
\ln A & = &  a_2 +b_2 \ln(\frac{\alpha}{\ln \alpha}),
\end{eqnarray*}
and for $\hat{q}$ and $\hat{R}$
\begin{eqnarray*}
A & = &  a_1 +b_1 \ln \alpha,\\
A & = &  a_2 +b_2 \ln(\frac{\alpha}{\ln \alpha}).
\end{eqnarray*}
In the Table I, we give the list of $a_i$ and $b_i$ for these quantities.
  In particular, we note that
  $T_c \rightarrow \infty$ as $\alpha \rightarrow \infty$.
Further, we obtained $\frac{\Delta q}{\Delta R}\sim 1.7$
and $\frac{\hat{R}}{\hat{q}}\sim 2.5$.
As an example, we show the asymptotic behaviour of 
$\Delta \epsilon_g$ in Figure 7.\\

\begin{table}
\caption{Coefficient $a_i$ and expontent $b_i$ extimated
for $200 \le \alpha \le 381$ and theoretical value $b_{2,th}$.}
%\label{table:1}
\begin{tabular}{@{\hspace{\tabcolsep}\extracolsep{\fill}}ccccccc} \hline
      & $\Delta q$ & $\Delta R$ & $T_c$  & $\Delta \epsilon_g$
& $\hat{R}$ & $\hat{q}$ \\
$a_1$ & 2.5 & 2.2 & -1.8 & 1.1&  -0.49 & 0.89 \\
$b_1$ & -1.5 & -1.5 & 0.81 & -1.5& 1.1 & 0.28 \\
$a_2$ & 1.1 & 0.88 & -1.1 & -0.27 & 0.50 & 1.1 \\
$b_2$ & -1.8 & -1.9 & 0.99 & -1.9 & 1.4 & 0.35 \\
$b_{2,th} $ & -2 & -2 & 1 & -2 & 1 & 1 \\ \hline
\end{tabular}
\end{table}

\noindent
(II) $\delta=0$

We treated the same function $P(y)$ as in the calculation for the
RS solution, $P(y)= \frac{1}{2} $ sgn $(y)$.
We depict $\alpha$ dependence of $T_c$ in Figure 8.
See Figure 16 for $\alpha$ dependence of $ q_0, R$ and $\Delta \epsilon _g$.
  In the 1RSB solution, there exist two branches I and II.  In the branch I
the quantities agree to those in the RS solution with $T=0$ as 
$\alpha \rightarrow \alpha_s(T=0)$.  On the other hand,
$q$ and $R$ tend to 1 as $\alpha$ tends to 0 in the branch II.\par
Within our calculation, it is difficult to determine
which case of (a)$0 \llsim \mu <1$ or (b) $\mu \ge 1$
takes place in the branch II,
since we could obtain solutions only for $\alpha \gsim 0.45$.
As for $T_c$ it seems that $T_c \rightarrow $ finite
as $\alpha \rightarrow 0$.  
In both branches, $\Lambda_1$ and $\Lambda_3$ are negative,
that is, AT stable.
As for the free energy,
% $f_{1RSB}(\beta) =  f_{RS}(\beta _c)$.  
$f^{I}_{RSB} < f^{II}_{RSB}$ holds.  See Figure 9.

\subsection{Asymptotic behaviors}
As is suggested from the above numerical results and will be shown later,
asymptotic behaviors of $\beta_{c}$ are different 
in  the cases of $\delta=0$ and $\delta>0$.  
Thus, we discuss these cases separately.\\

\noindent
(I) The case of $\delta>0$ \par
Suggested by numerical results we consider the limit 
$\beta_c \ll 1$.    For $\beta \ll 1$, $f_{RS}$ and $S_{RS}$ 
are estimated in the asymptotic region as  follows.  See Appendix E.
\begin{eqnarray}
  - \beta f_{RS} & = &
  -\frac{\hat{q}}{2} (1- q) - R \hat{R} + I \nonumber \\
&&  - \alpha \beta [ \epsilon_{min} +
\frac{2s}{(1+\delta) \sqrt{2\pi}}
(2 \Delta R)^{\frac{1+\delta}{2}}
- \frac{ \beta \sqrt{\Delta q}}{2 \pi \sqrt{2} } ], \\
 S_{RS}& =&  -\frac{\hat{q}}{2} \Delta q
- \hat{R} R + I
- \alpha \beta ^2 \frac{\sqrt{\Delta q}}{2 \pi \sqrt{2} }.
\end{eqnarray}
First, let us show that for $0 \llsim \mu <1$, 
no consistent 1RSB solution exists
in the present situation.
The saddle point equations are
\begin{eqnarray}
\Delta q & \simeq & \frac{2 h(\tau) \psi(\mu)}{ \sqrt{ \hat{q_c}}}, 
\label{eq:dqdlt}\\
\Delta R & \simeq & \frac{\Delta q}{2 \mu} ,\label{eq:drdlt}\\
\hat{q}_c &=& \hat{q}_0 \alpha
 \beta_c ^2 /\sqrt{\Delta q}, \label{eq:hqc}\\
\hat{R}_c &=& \hat{R}_0 \alpha
 \beta_c (\Delta R)^{(\delta -1)/2}\label{eq:hrc}, \\
\hat{q}_0 &=& \frac{1}{2 \pi \sqrt{2}}, \;\;\;
\hat{R}_0 =\frac{s}{\sqrt{\pi}} 2^{\delta/2}, \nonumber
\end{eqnarray}
where $\Delta q=1-q_c$ and $ \Delta R = 1 - R_c$.
Then, the condition that the entropy is zero becomes
\begin{equation}
\Delta q  \simeq  \frac{4 \psi(\mu) h(\tau) }{\hat{R} \tau}.
\label{eq:dqdlt2}
\end{equation}
From equations $(\ref{eq:dqdlt})$ and $(\ref{eq:dqdlt2})$, we obtain
$ \tau ^2 =2. $ 
Since we consider the case of $R \simeq 1$ and $q \simeq 1$,
$\tau$ should be large.  Thus, this case is inadequate.
 Therefore, in the below,
we consider the case of $1 \le \mu$.
  In this case,
$I \simeq \hat{R} + a_{\mu} e ^{-2(\hat{R} - \hat{q})}$.
Then, $f_{RS}$ and $S_{RS}$ become
\begin{eqnarray*}
-\beta_c f_{RS}& \simeq & -\frac{\hat{q}_c}{2} \Delta q
 + \hat{R}_c \Delta R  + a_{\mu} e^{-2(\hat{R}_c-\hat{q}_c)} 
  - \alpha  \beta _c
( \epsilon_g - \frac{\beta _c}{2 \pi \sqrt{2}} 
\sqrt{\Delta q} ), \\
S_{RS} & \simeq & -\frac{\hat{q}_c}{2}
 \Delta q +  \hat{R}_c \Delta R
- \frac{\alpha \beta_c ^2 }{2 \pi \sqrt{2}}\sqrt{\Delta q}
 + a_{\mu} e^{-2(\hat{R}_c-\hat{q}_c)}.
\end{eqnarray*}
The saddle point equations for $\hat{q}$ and $\hat{R}$ are
the same as in the case (a), and those for $q$, $R$ and
zero entropy condition are
\begin{eqnarray}
\Delta R &=& 2 a_{\mu}  e ^{-2(\hat{R_c} - \hat{q_c})}, \label{eq:drdlt2}\\
\Delta q &=& 2 \Delta R,\label{eq:dqdlt3}\\
S_{RS} & \simeq & -\frac{\hat{q}_c}{2} \Delta q +  \hat{R}_c \Delta R
- \alpha \beta_c ^2 \hat{q}_0 \sqrt{\Delta q}
+ a_{\mu} e^{-2(\hat{R}_c-\hat{q}_c)} = 0.\label{eq:ent}
\end{eqnarray}
Since $\hat{R}_c \gg 1$ and $\hat{q}_c \gg 1$,
using equations ($\ref{eq:hqc}$), ($\ref{eq:drdlt2}$)  and 
($\ref{eq:dqdlt3}$) , we obtain from equation ($\ref{eq:ent}$) 
the following relation,
\begin{equation}
\hat{R}_c = 3\hat{q}_c.\label{eq:rel1}
\end{equation}
That is, $ \mu=3/2$ and $a_{\mu}=1$.
Thus, from equations ($\ref{eq:hqc}$), ($\ref{eq:hrc}$) and 
 ($\ref{eq:rel1}$),
we obtain
\[
\hat{q}_c = F_0 \alpha e^{-2(2 \delta -1) \hat{q}_c},
\]
that is,
\[
 \ln \hat{q}_c = \ln \alpha -2( 2 \delta -1) \hat{q}_c + \ln F_0,
\]
where $F_0= 4 s^2 2^{2\delta}/9\sqrt{2}$.  
Thus, 
\[ \hat{q}_c \simeq \frac{ \ln \alpha }{2(2 \delta -1)} \]
for $2\delta-1 \ne 0$.  This implies that 
$\hat{q}_c$ tends to infinity
under the condition that
 $\alpha$ tends to infinity for $\delta >1/2$  or
$\alpha$ tends to 0 for $\delta <1/2$.
For $\delta = 1/2$, $\hat{q}_c =F_0 \alpha $.
Therefore, we obtain the following results.
\begin{enumerate}
\item In the case of $\delta > \frac{1}{2}$, as $\alpha \rightarrow \infty$,
\begin{eqnarray}
&&\Delta R  \simeq 2 ( \frac{\ln \alpha}{  \alpha})
^{\frac{2}{2\delta-1}}, \;\;
\Delta q  \simeq 2  \Delta R, \\
&& \hat{R}_c \simeq \frac{3}{2(2\delta-1)}
 \ln (\frac{ \alpha}{ \ln  \alpha}),
\;\; \hat{q}_c \simeq \hat{R}_c /3 , \;\;
\beta_c  \simeq
\frac{4 \sqrt{\pi} s}{3} 2^{\delta}
 ( \frac{\ln \alpha}{  \alpha})
^{\frac{\delta}{2\delta-1}},\\
&& \Delta \epsilon _g \simeq \epsilon_0 (\Delta R)^{\frac{\delta+1}{2}}
\simeq \epsilon_0 2^{\frac{\delta+1}{2}} 
 ( \frac{\ln \alpha}{  \alpha})^{\frac{\delta+1}{2\delta -1}},\;\;
 \epsilon_0 = \frac{2s}{(1+\delta)\sqrt{2\pi}} 2^{\frac{1+\delta}{2}}.
\end{eqnarray}
\item
In the case of $\delta = \frac{1}{2}$, as $\alpha \rightarrow \infty$,
\begin{eqnarray}
&& \Delta R \simeq 2 e^{-4F_0 \alpha}, \;\;
\Delta q  \simeq 2 \Delta R, \\
&& \hat{R}_c \simeq 3  F_0 \alpha,\;\; \hat{q}_c \simeq \hat{R}_c/3,\;\;
\beta_c \simeq  \frac{4 s}{3} \sqrt{2 \pi}
e^{-F_0 \alpha},\\
&& \Delta \epsilon_g \simeq \epsilon_0 2^{\frac{3}{4}} e^{-3F_0 \alpha}.
\end{eqnarray}
\item
In the case of $0 < \delta < \frac{1}{2}$, 
as $\alpha \rightarrow 0$,
\begin{eqnarray}
&&\Delta R \simeq
2  ( \frac{ \alpha}{ \ln \frac{1}{ \alpha} })
^{\frac{2}{1 - 2\delta}}, \;\;
\Delta q  \simeq 2  \Delta R, \\
&& \hat{R}_c \simeq \frac{3}{2(1-2\delta)}
\ln (\frac{1}{ \alpha} \ln \frac{1}{\alpha}  ), \;\;
 \hat{q}_c  \simeq  \hat{R}_c/3, \;\;
\beta_c \simeq   \frac{4 \sqrt{\pi} s}{3} 2^{\delta}
 ( \frac{ \alpha}{ \ln \frac{1}{ \alpha} })
^{\frac{\delta}{1 - 2\delta}},\\
&& \Delta \epsilon_g \simeq \epsilon_0 2^{\frac{\delta+1}{2}} 
 ( \frac{\alpha}{\ln \frac{1}{\alpha}})^{\frac{1+\delta}{1-2\delta}}.
\end{eqnarray}
\end{enumerate} 

Thus, when $0< \delta <1/2$, for large $\alpha$, there is no solution
such that $q_c \rightarrow 1$ and $R_c \rightarrow 1$.
This implies that there is a value of $\alpha=\alpha_{max}$
such that for $\alpha > \alpha_{max}$ there is no solution
except for the PL when $0< \delta <1/2$.

Here, let us compare the theoretical results with numerical ones
on asymptotic behaviours for $\delta =1$.  As is shown in
the Table I, we note that numerically obtained exponents
 $b_2$ and theoretical ones $b_{2, th}$
 agree fairly well except for those of $\hat{q}$ and $\hat{R}$.
$\hat{q}$ and $\hat{R}$ are proportional to $\ln(\frac{\alpha}{\ln \alpha})$
and it is difficult to estimate a logarithmic dependence directly.
Instead, we can check the theoretical result of the
relation $\frac{\hat{R}}{\hat{q}}=3$.
Numerically, this value is 2.5.  Further, as for the relation
$\frac{\Delta{R}}{\Delta{q}}=2$, we obtained numerically 1.7.
Therefore, we can conclude that 
the agreement between theoretical and numerical results are fairly well.

Now, let us examine the case $\delta=0$.
In this case, substituting $\delta=0$ 
into the above expressions for the case of $0<\delta<1/2$,
$\beta_c$ becomes constant.  
That is, the assumption $\beta_c \ll 1$ is not 
satisfied.  This is the reason why we treat the cases 
$\delta>0$ and $\delta=0$ separately.\\
\noindent
(II) The case of $\delta=0$.\par
In \S3, we examined the RS solution  
with $\beta $ fixed.  Now, let us investigate the 1RSB solution 
by imposing the condition $S_{RS}=0$.\par
(a)  $0 \llsim \mu < 1 $ \par
In this case, $ S_{RS}$ becomes
\[ S_{RS} \simeq \frac{\hat{q} \Delta q}{2} 
+ \alpha \sqrt{\Delta q}\; r.\]
Then, from the condition $ S_{RS}=0 $ we obtain
\begin{equation}
%\[ \hat{q}= -2 \alpha \frac{r_4}{\Delta q}.\]
 g_{1, 0}(\chi, \beta_c) = -2 r(\chi, \beta_c). 
\label{eq:consa}
\end{equation}
From equations $(\ref{eq:rschia})$ and $(\ref{eq:consa})$,
$\chi$ and $\beta_c$ are determined.
The 1RSB solution appears for $\beta > \beta_c$.\par
(b) $\mu \ge 1 $ \par
$S_{RS}$ is 
\[ S_{RS}\simeq -\frac{\hat{q} \Delta q}{2}
+ \frac{\hat{R} \Delta q}{2} 
+ \alpha \sqrt{\Delta q}\; r. \]
Then, the condition  $S_{RS}=0$ becomes
\begin{equation}
g_{1, 0}(\chi=0,\beta _c)=g_{2, 0}
(\chi=0,\beta _c)
+2r(\chi=0,\beta _c).\label{eq:consb}
\end{equation}
If the solution $\beta_c$ of equation $(\ref{eq:consb})$ exists, 
1RSB solution appears for $\beta > \beta_c$.
The condition of the existence of this
type of solution is $ g_{2, 0} / g_{1, 0} \ge 2$.\\
Numerical calculations for $\delta=0$ indicate that $\chi$ 
tends to constant 
and then the case (a) appears.
Thus, for $\alpha \ll 1$
\begin{eqnarray}
\Delta R & \simeq & \frac{1}{ \sqrt{2 \pi}} \psi(\mu_c)
\alpha ^2 ( \ln \frac{1}{ \alpha} )^{-2},
 \; \; \Delta q=2\mu_c \Delta R,\\
\hat{R}_c & \simeq & \frac{2}{ \mu_c}
\ln [ \frac{1}{\alpha } ( \ln \frac{1}{\alpha })^{3/4} ],
\; \; \hat{q}_c =\frac{\hat{R}_c}{2\mu _c},\\
\Delta \epsilon_g  &= & \epsilon _0 \sqrt{\frac{\psi(\mu_c)}{2\pi}}
\alpha  ( \ln \frac{1}{ \alpha} )^{-1}.
\end{eqnarray}
This implies that for $\delta =0$, the PL takes place.

As for the validity of the above asymptotic solutions for any $\delta \ge 0$,
since the coefficient of any quantity 
 does not contain $\beta$, there is no condition
for the range of $\beta$.

Thus, in the case of $0 \le \delta < \frac{1}{2}$,
there exists no solution for $\alpha > \alpha_{max}$.
This is consistent with 
the result derived in \S4
 that the PL exists for $0 \le \delta<\frac{1}{2}$ when 
$0<\beta < \infty$.\vspace{5mm}\\

Putting together all results obtained in this paper,
we get the following behaviours of learning.

When  $T$ is small enough, there is a critical value of 
$\alpha$, $\alpha_s(T)$ above which
the entropy of the RS solution becomes negative.
Thus, for $\alpha > \alpha_s(T)$, the 1RSB solution appears.
Within the 1RSB ansatz, we found that 
the behaviour of the generalization error $\epsilon_g$ 
is classified into the following 
three categories
according to the value of $\delta$.
\setcounter{enumi}{3}
\begin{enumerate}
\item If $0 \le \delta<\frac{1}{2}$, 
solutions with $R<1$ exist only for a finite range of $\alpha$,
$[0, \alpha_{max}]$.  
There is a critical temperature $T_c$.
When $T>T_c$, the entropy of the RS solution is positive and
this solution is AT stable. 
When $T<T_c$, for $\alpha >\alpha_s(T)$, the 1RSB solution appears.
In both cases, at $\alpha = \alpha_{max}$, 
 a first-order phase transition
from the RS solution with positive entropy or from the 1RSB solution to 
the Perfect Learning takes place.
\item If $\delta=\frac{1}{2}$,
$\alpha_s(T)$ is defined for any temperature $T$,
and 1RSB solution appears
for $\alpha>\alpha_s(T)$.
$\epsilon_g$ for the 1RSB solution decays exponentially, 
\[\Delta \epsilon_g \sim e^{- 3F_0 \alpha},\]
where $F_0$ is a constant.
\item If $\delta > \frac{1}{2}$, 
for any temperature $T$,
$\alpha_s(T)$ is defined and 1RSB solution appears
for $\alpha>\alpha_s(T)$.
$\epsilon_g$ for the 1RSB solution decays as a power law
with a logarithmic correction, 
\[ \Delta \epsilon_g \propto ( \frac{\ln \alpha }{\alpha})
^{\frac{1+\delta}{2\delta-1}}. \] 
\end{enumerate}
To check these theoretical results,
we performed numerical calculations.
In the next section, we give the results of the calculations. 

\section{Numerical Calculations by exhaustive method}
We performed numerical calculations  by the exhaustive method
for $\delta=0$ and $\delta=1$.  
We used the minimum-error algorithm
and the Gibbs algorithm for several temperatures.  We calculated 
several quautities such as $q, R, \epsilon_g$, etc.
For example, $q$ and its standard deviation $\delta q$
are calculated by the following formulas.
\begin{eqnarray}
 q& =& \frac{1}{N_{\xi}} \sum_{\xi} q_{\xi}, \;\;
q_{\xi}= \sum_{\alpha < \beta} q^{\alpha,\beta}P_{\alpha} P_{\beta}
/ \sum_{\alpha < \beta}P_{\alpha} P_{\beta}, \label{eq:formula1}\\
\delta q & =& \sqrt{ [ \sum_{\xi}q_{\xi} ^2 - 
(\sum_{\xi}q_{\xi})^2/N_{\xi}]/(N_{\xi}-1)},\\
&& P_{\alpha}=e^{- \beta E_{\alpha}}/\sum_{\alpha} e^{-\beta E_{\alpha}},
\end{eqnarray}
where $\alpha$ denotes one of $2^N$ configurations of
weight vectors and $E_{\alpha}$ is its energy,
$q_{\xi}$ is the thermal average for a given example $\xi$
and $N_{\xi}$ is the number of samples.
The calculations were performed for  $N$ up to 20 with
$N_{\xi}=200$.  First, let us show the results for $\delta=1$.\\
\noindent
(I) $\delta=1$

The $i$-th component of an example $x_i$ is corrupted
by a gaussian noise $\eta _i$ with mean 0 and standard deviation 1.
This corresponds to $P(y)=1-2 H(y)$.  
First, we show the results for
the minimum-error algorithm.  
  In Figure 10, to see the system size ($N$) dependence of quantities, 
we show the $\alpha$ dependence of $R$ and its standard deviation $\delta R$
for $N=10, 15$ and 17.
From these results, it seems that the calculations in 
$N=15$ is sufficient  to obtain $N=\infty$  results 
at least for $\alpha $ up to 15.
In Figure 11, we compare numerical results with theoretical ones.
The agreement between the theoretical and numerical results
is fairly well except for $q_0$.
As for $q_0$,  we calculate it in each sample by the formula
 $(\ref{eq:formula1})$. 
In general, $q_0$ exhibits a large finite size effect,
because it is calculated for pair of states.
Thus, when the number of states with the minimum energy becomes small,
the fluctuation of $q_0$ becomes very large.
This is the reason why the agreement 
between the numerical and the theoritical results for
$q_0$ is worse than other
quantities.  A suitable quantity is the distribution
of $q$, $P(q)$.  We calculate this in the case of the Gibbs Algorithm.
In Figure 12 we show the $\alpha$ dependence of $R$ for larger value of
$\alpha$ in the case of $N=15$. At $\alpha \gsim 85$, there exists
only one state.  Since theoretically $R$ tends to 1 as $\alpha$
goes to $\infty$, to see whether this is a finite size effect or not
 we estimated the value of $\alpha_{max}$ 
 by the condition that $R$ exceeds $1-1/N$ for the first time as
$\alpha$ increases.  Then, we found that as $N $ increases 
$\alpha_{max}$ increases.  Thus, it seems that we observed the
finite size effect.

Next, we show the results for the Gibbs algorithm.  In this algorithm,
we calculated for $N=10$ and 12 and for several
temperatures, and we took into account the all staes.
We confirmed that the results for $N=10$ and 12 are almost same.
We show both numerical and theoretical results
in Figure 13 for $R$ and $\Delta \epsilon_g$
and in Figure 14 for $q$ and $q_0$.
The distribution of $q$, $P(q)$, is also shown for
several temperatures together with theoretical results in Figure 15.
$P(q)$ is calculated by
\[
 P(q)=<\sum_{\alpha,\beta} \delta(q,q^{\alpha \beta})P_{\alpha} P_{\beta}>,
\]
where $\delta(q,q^{\alpha \beta})$ is the Kronecker's delta and
$<\cdot>$ means the sample average.
From these, we see the agreement between theoretical 
and numerical results is fairly well. \par

\noindent
(II) $\delta=0$

The output by a teacher is reversed with the probability $(1-k)/2$
with $k=\frac{1}{2}$, that is we treat the same case as before.
First, we show the results for the
minimum-error algorithm.  
We investigated the $N$ dependence of $R(\alpha)$ and
$\delta R(\alpha)$ for $N=10, 15, 17$ and 20 
and found that the results for $N=15, 17$ and 20 are almost same.
In Figure 16, $\alpha$ dependence of several quantities are depicted
  for $N=17$ together with theroretical results.  
In the figure of $q$ and $q_0$(Figure 16(b)), we note that 
$q_0$ takes values at above the theoretical upper bound of $\alpha$,
$\alpha_{max}$. This is due to a finite size effect
mentioned in the above.
In Figure 17 we show the behaviour of $R$ for larger value of
$\alpha$ in the case of $N=15$. For $\alpha>22$, there exists
only one state.  In the  case of $N=10$, it occurs for 
$\alpha>25$.  To investigate whether the PL exists or not,
we numerically estimated $\alpha_{max}$ by the same method as
that in the case of $\delta=1$.  Contrary to the case of $\delta=1$,
$\alpha_{max}$ decreases as $N$ increases.  
Thus, we conclude the PL takes place
even when $N=\infty$.   Theoretically, 
$\alpha_{max}$ is about 9.13.

Next, we show the results for the Gibbs algorithm.  In this algorithm,
we calculated for $N=10$ and 12, and 
for several temperatures.
We took into account the all staes.
We confirmed that $N=12$ is sufficient for convergence for any $T$.
In Figure 18, $\alpha$ dependence of $R$ and $\Delta \epsilon _g $
are shown for $T= 1.0 $.
Also, $\alpha$ dependence of $q$ and $q_0$ 
are depicted in Figure 19 for 
 $T= 0.15, 0.5 $ and 5.0.
As is shown in these figures, the agreement between 
the theoretical values and numecical ones are fairly well
except for the case of $q_0$.
Concerning $q_0$, we further calculated $P(q)$ at $\alpha =5$  for
several temperatures.  In Figure 20, we show the numerical results of
$P(q)$ together with theoretical ones.
The positions of peak values agree in the theoretical and numerical results.

In conclusion, in both cases of $\delta =1$ and 0, 
although there exist finite size effects
 as is seen for $q_0$, 
as a whole theoretical results and numerical ones 
agree fairly well within the 1RSB ansats.

\section{Summary and Discussion}
In this paper, we studied the learning from stochastic examples
by perceptrons with Ising weights.
By using the replica method, we obtained the condition for
the existence of the Perfect learning and
power law of learning curves in the asymptotic region,
in terms of $\delta$ which
represents the local property of the rules by which examples are drawn.
First, let us summarize the results in more details.

Our assumptions are as follows.\\
When an input vector $\vecx$ is given,
the probability $ p_r(+1| \vecx) $ that a teacher returns an
output $+1$  is a function
of the inner product between the input $\vecx$ and the teacher's
weight $\veco$ and take the following form,
\begin{eqnarray*}
  p_r(+1| \vecx)  & = & {\cal P}(u^o)=\frac{1+P(u^o)}{2}, \\
  u^o & \equiv &
( \vecx \cdot \veco)/\sqrt{N},\;\; |\vecx|= \sqrt{N},
\;\;|\veco|= \sqrt{N}.
\end{eqnarray*}
Further, we assume $P(y)$ is non-decreasing and near $y = 0$
it behaves as 
$ P(y) \simeq a \; {\rm sgn}(y) |y|^{\delta},
 \; \; (\delta \ge 0)$.  For simplicity, we assumed that $ P(y)$ 
is an odd function.\\
Under the above assumptions, we obtained the following results.

\noindent
{\bf Conditions for the PL}

The the necessary and sufficient conditions for the
existence of the Perfect Learning are

(a)$ 0\le \delta <1/2$ \mbox{ when }$0<\beta <\infty$,

(b)  deterministic case.

\noindent
{\bf Behaviour of learning curves}

Within the 1RSB ansatz, we found that 
the behaviour of the generalization error $\epsilon_g$ 
is classified into the following 
three categories
according to the value of $\delta$.
\setcounter{enumi}{3}
\begin{enumerate}
\item If $0 \le \delta<\frac{1}{2}$, 
at $\alpha = \alpha_{max}$, 
 a first-order phase transition
from the RS solution with positive entropy or from the 1RSB solution to 
the Perfect Learning takes place.
\item If $\delta=\frac{1}{2}$,
for large $\alpha$ 1RSB solution appears
and $\epsilon_g$ for the 1RSB solution decays exponentially, 
\[\Delta \epsilon_g \sim e^{- 3F_0 \alpha},\]
where $F_0$ is a constant.
\item If $\delta > \frac{1}{2}$, 
for large $\alpha$ 1RSB solution appears
and $\epsilon_g$ for the 1RSB solution decays as a power law
with a logarithmic correction, 
\[ \Delta \epsilon_g \propto ( \frac{\ln \alpha }{\alpha})
^{\frac{1+\delta}{2\delta-1}}. \] 
\end{enumerate}

To check these results, we performed several numerical
calculations.  Those are, the calculations of the saddle point 
equations for the RS and the 1RSB solutions and the
direct calculations of concerned quantities by enumeration methods.
The numerical and theoretical results showed fairly well agreement.
As is mentioned in the introduction,
Seung also investigated 
the existence of the Perfect Learning
when the weights are Ising and
a rule to be learnt is stochastic \cite{Seung 95}
by the annealed approximation.
He classified learning
behaviour of Ising networks
by introducing the following two exponents $y$ and $z$.
The first exponent $y$ is associated with  
$\rho (\epsilon_g)$
which is the logarithm of the number of weight
 vectors whose 
generalization errors have a value $\epsilon_g$.
He assumed that when $\Delta \epsilon_g= \epsilon_g-\epsilon_{min}$
 is small, $\rho (\epsilon_g)$ increases as
$ \rho (\epsilon_g) \sim {\cal O }
((\Delta \epsilon_g)^y)$,
where $\epsilon_{min}$ is the minimum value of the 
generalization error obtained by the unique
 optimal weight vector $\veco$.
The second exponent $z$ is introduced to characterize
$e_d(\vecw, \veco)$ which is the probability 
that the output
for the weight vector $\vecw$ differs from that 
for the optimal weight vector $\veco$.
He also assumed that $e_d(\vecw, \veco)$ is scaled as
$ e_d(\vecw, \veco) \sim {\cal O } 
((\Delta\epsilon_g)^z)$.
He estimated the upper bounds for the generalization errors
and found that the behavior of learning curves varies
according to the values of indices $y$ and $z$.
His results are summarized as follows.
\begin{enumerate}
\item  If $y+z>2$, there is a first-order transition.
\item  If $y+z<2$, the generalization error decays as a
power law,  $\Delta\epsilon_g \sim
\alpha ^{-\frac{1}{2-y-z}}$.
\item  If $y+z=2$, there is a second-order transition
 or the generalization error decays exponentially.
\end{enumerate}
In our model, the exponents $y$ and $z$ are expressed as
$ y= \frac{2}{1+ \delta}, \; \;
 z= \frac{1}{1+ \delta} = \frac{y}{2}$
respectively, and then $ v =  \frac{3}{1+ \delta}.$
Therefore, $ \delta = \frac{3-v}{v} $
follows and it is found that our results on the typical
 learning behavior agrees with Seung's results
which are the upper bounds of the learning curves.\par

As for the condition of the existence of the PL, we note that
 for $\beta = \infty$, i.e. $T=0$, 
the PL does not exist in the learning from 
stochastic examples.
The reason is that for $T=0$ and for large $\alpha$
there exists no student whose outputs are the same
as the teacher's, since the
teacher makes mistakes.  Thus, the volume of 
weight vectors 
whose energies are 0 vanishes for large $\alpha$.
On the other hand, for the case of 
$T \rightarrow +0$, we consider
the weight vectors of the minimum energy, and there is 
at least one solution of $\vecw=\veco$
   when $\alpha$ is large enough.  Thus, the PL is possible
in the limit $T \rightarrow 0$.\par
As the learning advances, $\vecw$ tends to $\veco$.  
The examples which give the crucial
influence on the learning are such that 
$u_0 = ( \vecx \cdot \veco )/
\sqrt{N} \sim 0$.  The more slowly 
the probability ${\cal P}(u)$
varies around $u=0$ for larger $\delta$, 
the more difficult students tune the optimal vector
$\veco$.  This is the reason why the larger $\delta$  is,
 the Perfect Learning becomes 
more difficult to take place.

\section*{Acknowledgements}
The authors is grateful to Y. Kabashima 
 for valuable discussions and collaborations in the first part
of this study.  He is also grateful to A. C. C. Coolen and
P. Sollich for valuable discussions and suggestions
during his stay in King's College, University of London.

\appendix
\section{Derivation of free energy}
Here, we derive the free energy by the
replica method.
Introducing $n$ replicas, the partition function $Z^n$ becomes
\begin{eqnarray}
Z^n  &= & \mbox{Tr} \prod _{\alpha =1} ^n 
[ e^{-\beta} \int_{-\infty}^0 d\lambda^{\alpha} _{\mu}
 + \int_0^{\infty}d\lambda^{\alpha} _{\mu} ]
\int _{-\infty} ^{\infty}  \frac{dy_{\mu} ^{\alpha}}{2\pi}
\exp[ -iy_{\mu}^{\alpha}(r^o_{\mu}  u_{\mu} ^{\alpha}
 - \lambda_{\mu} ^{\alpha})],
\end{eqnarray}
where $u_{\mu} ^{\alpha}=( \vecx ^{\mu} \cdot  \vecw
^{\alpha})/\sqrt{N}$
and Tr implies the summation over all 
configurations of $\vecw^{\alpha}, \alpha=1, \cdots n.$
Defining 
 the overlap between the weight vector of a learner and the optimal weight
 vector, 
 $R ^{\alpha} =  \frac{1}{N} \sum_{ j=1} ^{N} w_{j}^{\alpha} w_{j}^{o}$
, and the overlap between the weight vectors of learners,
$q ^{\alpha \beta} =  \frac{1}{N}
\sum_{ j=1} ^{N} w_{j}^{\alpha} w_{j}^{\beta}$,
and using the relations
\begin{eqnarray*}
&&   \delta( \sum_{j=1}^{N}(x_j ^{\mu})^2 -N)
  =  \int_{-i \infty}^{i \infty}
 \frac{d \tilde{K_{\mu}}}{2\pi i} \exp
  [ -\tilde{K_{\mu}} 
( \sum_{j=1}^{N}(x ^{\mu}_{j})^2 -N) ], \\
&&1=  \prod_{\alpha} \int dR^{\alpha}
\int_{-i \infty}^{ i\infty}
 \frac{Nd \hat{R}^{\alpha} }{2\pi i} \exp [
 - N \hat{R}^{\alpha} ( R^{\alpha} -
\frac{1}{N}\sum_{j=1}^{N}w_{j}^{\alpha}w_{j}^o ) ], \\
&&1 = \int \prod_{\alpha < \beta} dq^{\alpha \beta}
\int_{-i \infty}^{i \infty}
 \frac{Nd \hat{q}^{\alpha \beta} }{2\pi i} \exp
[ - N \hat{q}^{\alpha \beta}( q^{\alpha \beta} -
\frac{1}{N}\sum_{j=1}^{N}w_{j}^{\alpha}w_{j}^{\beta}) ], 
\end{eqnarray*}
we take the average over $r_{\mu} ^o$, 
and $\vecx ^{\mu}$, and obtain
  the expression for $<Z^{n}>_{\xi _p, w^o}$,
\begin{eqnarray}
\hspace*{-20mm}
 <Z^{n}>_{\xi _p, w^o}  & = &
  \int [ \prod_{\alpha< \beta} 
 dq^{\alpha \beta}\frac{Nd \hat{q}^{\alpha \beta} }{2\pi i}]
[\prod_{\alpha} dR^{\alpha} \frac{Nd \hat{R}^{\alpha} }{2\pi i}] e^{NG} , \\
\hspace*{-20mm}
G &=& \frac{p}{N} G_1(\{q^{\alpha \beta}\},\{ R^{\alpha}\})
    + G_2(\{\hat{q}^{\alpha \beta}\},\{ \hat{R}^{\alpha}\})
  -\sum_{\alpha} \hat{R^{\alpha}}R^{\alpha} 
  - \sum_{\alpha<\beta} \hat{q}^{\alpha\beta}q^{\alpha\beta},\\
\hspace*{-20mm}
 e^{G_1} &=&  [ \prod_{\alpha} 
 ( e^{-\beta} \int_{-\infty}^0 d\lambda^{\alpha}
 + \int_0^{\infty}d\lambda^{\alpha} )
 \int _{-\infty} ^{\infty}  \frac{dy ^{\alpha}}{2\pi} ] \nonumber \\
\hspace*{-20mm}
&& \times \exp[ - \frac{1}{2} \sum_{\alpha} (y^\alpha)^2 
- \sum_{\alpha<\beta} q^{\alpha \beta}y^{\alpha} y^{\beta} 
  + i\sum_{\alpha} y^\alpha \lambda^{\alpha} ]
\Psi ( \sum_{\alpha} y^{\alpha} R^{\alpha}),  \\
\hspace*{-20mm}
   e^{G_2}& =& \mbox{Tr}
\exp[ \sum_{\alpha} \hat{R} ^{\alpha} w ^{\alpha}
+ \sum_{\alpha < \beta} \hat{q}^{\alpha \beta}
 w^{\alpha}w^{\beta}],\\
\hspace*{-20mm}
&& \Psi(y) \equiv  \frac{1}{ \sqrt{2 \pi}} \int d \xi 
e^{ - \frac{1}{2}(\xi -i y)^2} \{1 - \frac{1}{2}[ P(\xi) - P(-\xi) ]\}.
 \nonumber 
\end{eqnarray}
where Tr implies the summation over $w^{\alpha}, \alpha =1, \cdots n$.
In the above expressions we set $\tilde{K}_{\mu}=1/2$ 
which is the optimal value.  When $P(-y)=-P(y)$, $\Psi(y)$ 
becomes 
\begin{equation}
\Psi(y)=\frac{1}{ \sqrt{2 \pi}} \int d \xi
e^{ - \frac{1}{2}(\xi -i y)^2} 2 {\cal P}(-\xi).
\end{equation}
The general form of the free energy per weight is given by
\begin{equation}
 f= - \frac{ < \ln  Z >_{\xi_p, w^o}} {N \beta}
   = -  \frac{G} {n \beta}.
\end{equation}
\vspace{0.5cm}

\section{ Derivation of asymptotic relations for $\hat{q}$ and $\hat{R}$
when $q \rightarrow 1$ and  $R \rightarrow 1$} 
In this appendix, we briefly derive
the asymptotic relations for $\hat{q}$ and $\hat{R}$.

First, let us consider the case of $0<\beta<\infty$.
The equation $(\ref{eq:s3})$ for $\hat{q}$ is rewritten as follows.
\begin{eqnarray}
\hat{q}& =& \frac{\alpha Q}{1-q}\frac{(1- e ^{-\beta})^2}{\sqrt{2 \pi}}
(A-B),\\
&& A = \int du e^{-Q^2 u^2/2}\frac{h(u)^2}{\tilde{H}(u) ^2},\\
&& B =\frac{1}{\sqrt{2+Q^2}}\frac{1}{\sqrt{2 \pi}}
 \int_0 ^{\infty} Dz P(\varepsilon z)\int_{-\infty} ^{\infty} Dt
H_2 [ \kappa(t+\frac{1}{\chi} \sqrt{\frac{1-\xi ^2}{2 + Q^2}}z)],
\label{eq:bb2}
\end{eqnarray}
where $H_2(u)=\frac{1} {\tilde{H} ( u)^2} -
\frac{1}{\tilde{H}(-u)^2} $,
$\kappa = \frac{\chi}{\sqrt{1+2\chi ^2}}$ and
$\varepsilon = \sqrt{\frac{Q^2 + 2 \xi ^2}{2+ Q ^2}}$.
It follows that $ H_2(u)$ is strictly increaseng odd function and 
$0< | H_2(u)| < e^{2\beta}-1 $ for $u \ne 0$.
Thus, for $\delta >0$, $P(\varepsilon z)$ can be replaced by
$a(\varepsilon z)^{\delta}$ in the equation (\ref{eq:bb2}).
\begin{eqnarray*}
\hspace{-10mm}
B & \simeq & \frac{1}{ 2 \sqrt{\pi}}
 \int_0 ^{1/\varepsilon} Dz a (\varepsilon z)^{\delta}
\int_{-\infty} ^{\infty} Dt
H_2 [\kappa(t+\frac{1}{ \sqrt{2} \chi} z)]
+{\cal O}( H(1/\varepsilon)). 
\end{eqnarray*}
As $q$ and $R$ tend to 1, $\varepsilon$ tends to 0 and then
 $B \rightarrow 0$.
On the other hand, for $T>0$ $A$ is finite for these limits.
Thus, $A-B \simeq A$.  Therefore, we obtain for $\delta >0$,
\begin{eqnarray}
\hat{q} & \simeq & \frac{\alpha}{ \sqrt{\Delta q}} 
g_{1, \delta}(\chi , \beta), \\
&& g_{1, \delta }(\chi , \beta)=
\frac{1}{\sqrt{2 \pi}}
 \int du \tilde{\varphi}(u)^2,
\end{eqnarray}
where $\Delta q \equiv 1 - q$. 
For the case of $\delta =0$, from equation $(\ref{eq:bb2})$, we obtain
\begin{eqnarray*}
\hspace{-10mm}
B & \simeq & \frac{1}{ 2 \sqrt{\pi}}
 \int_0 ^{\infty} Dz k
\int_{-\infty} ^{\infty} Dt
H_2 [ \frac{\chi}{\sqrt{1+2\chi ^2}} (t+\frac{1}{ \sqrt{2} \chi} z )] 
  = k \int du \frac{h(u)^2}{\tilde{H}(u) ^2}
[1-2H(u/\chi)],
\end{eqnarray*}
where $k=\lim_{y \rightarrow +0}P(y)$.
Then, we obtain 
\begin{eqnarray}
\hspace{-10mm}
\hat{q} & \simeq & \frac{\alpha}{\sqrt{\Delta q}} g_{1,0}(\chi, \beta),\\
&& g_{1,0}(\chi, \beta)
 = \frac{1}{\sqrt{2\pi}}  \int du \tilde{\varphi}(u)^2[ 1- k + 2k H(u/\chi)].
\end{eqnarray}
Now, let us estimate $\hat{R}$.
The equation $(\ref{eq:s4})$ is rewritten as
\begin{eqnarray}
\hspace{-20mm}
\hat{R}& =&  \frac{ \alpha}{q \xi}
\frac{(1- e^{-\beta})}{\sqrt{2\pi}}D,\\
\hspace{-20mm}
D&=& \int du e^{-Q^2 u^2/2}\frac{h(u)}{\tilde{H(u)} }w(u)
= \nu \int_{-\infty} ^{\infty} Dx H_1( \nu x) 
 \int_0 ^{\infty} Dy P(\xi y) e^{-\eta x y}(y+\eta x),\label{eq:d1}\\
\hspace{-20mm}
&& H_1( x)= \frac{1}{\tilde{H}( x)} + \frac{1}{\tilde{H}(- x)},
\;\; \nu= \frac{\xi}{\sqrt{\xi ^2 + Q^2}},\;\; 
\eta= Q \sqrt{ \frac{1-\xi^2}{\xi ^2 + Q^2}}.\nonumber
\end{eqnarray}
For $\delta>0$, $D$ is calculated as
\begin{eqnarray}
D&=& \frac{\xi}{\sqrt{1+Q^2}} \int_0 ^{\infty} Dy P'(\zeta z)
\psi(\frac{1}{\sqrt{1 + \chi ^2}} \sqrt{\frac{1- \xi ^2}{1 + Q^2}} z),\nonumber
\end{eqnarray}
where $\zeta=\sqrt{\frac{Q^2+\xi ^2}{1+Q^2}}$ and
 $\psi(z) =
 \int_{-\infty} ^{\infty} Dt H_1 (\nu  t -  z)$.
Since $\psi(z)$ is bounded, $D$ is evaluated as follows.
\begin{eqnarray*}
\hspace*{-10mm}
 D & \simeq & \xi [
\int_0 ^{1/\zeta}Dz a \delta (\zeta z) ^{\delta -1}
\psi(\frac{z}{\sqrt{1 + \chi ^2}}) 
+ {\cal O}\{H(1/ \zeta)\} ] 
 \simeq  \xi  a \delta \zeta ^{\delta -1}
\int_0 ^{\infty}Dz z ^{\delta -1}
\psi(\frac{z}{\sqrt{1 + \chi ^2}}). 
\end{eqnarray*}
That is, we obtain
 \begin{eqnarray}
\hspace{-10mm}
\hat{R} & \simeq & \alpha \frac{\xi ^{\delta}}{\sqrt{ \Delta q}}
 g_{2, \delta}(\chi , \beta),\\
\hspace{-10mm}
&&  g_{2, \delta}(\chi , \beta)  \equiv 
 \frac{a \delta }{\sqrt{2\pi}}
 (1 - e^{-\beta}) \frac{1}{\chi} (1+ \chi^{-2})^{(\delta -1)/2} 
 \int_0 ^{\infty}Dz z ^{\delta -1} 
\psi(\frac{z}{\sqrt{1 + \chi ^2}}).
\end{eqnarray}
For $\delta =0$, from equation $(\ref{eq:d1})$, we obtain
\[
D = \nu\int_{-\infty} ^{\infty} Dx H_1(\nu x)
[ \frac{k}{\sqrt{2 \pi}} 
+\xi \int_0 ^{\infty} Dy e^{-\eta x y} P'(\xi y) ].
\]
We assume that $|P'(y)|$ is bounded.
\footnote{
Here, we assumed the boundedness of $|P'(y)|$.
However, we can obtain the same results as those obtained here
without using $P'(y)$.}

  Then the
 second term in the parenthesis is evaluated as ${\cal O}(\xi)$
and is neglected.  Then,
\[
D \simeq \frac{k \nu}{\sqrt{2 \pi}} 
\int_{-\infty} ^{\infty} Dx H_1(\nu x)
= \frac{2k }{\sqrt{2 \pi}} \frac{\chi }{\sqrt{1+\chi ^2}}
 \int_{-\infty} ^{\infty} Dx
\frac{1}{\tilde{ H}(\frac{\chi x}{\sqrt{1+\chi ^2}})}.
\]
Thus, we obtain
 \begin{eqnarray}
\hat{R} & \simeq & \alpha \frac{1}{\sqrt{ \Delta q}}
g_{2, \delta},\\
&& g_{2, \delta} =
 \frac{k(1 - e^{-\beta})}{\pi }
 \frac{1}{\sqrt{1+ \chi^2}}
 \int_{-\infty} ^{\infty} Dx
\frac{1}{\tilde{ H}(\frac{\chi x}{\sqrt{1+\chi ^2}})}.
 \end{eqnarray}

Now, let us consider the case of $\beta= \infty$.  In this case, 
 $\tilde{\varphi}(u)$ becomes  $\varphi (u) \equiv \frac{H'(u)}{H(u)}$.
Therefore, from the equation (\ref{eq:s3}) we obtain
\begin{eqnarray}
\hspace*{-20mm}
\hat{q} & = & \frac{\alpha}{ \Delta q}
\int Du  \varphi(u/Q) ^2 E(u/Q) 
 \simeq  \frac{\alpha}{ \Delta q}(A'-B'),\nonumber\\
\hspace*{-20mm}
&& A'=\int_0 ^{\infty} Du  (u/Q)^2 = \frac{1}{2 Q^2},\nonumber\\
\hspace*{-20mm}
&& B'
 =\frac{\xi}{\sqrt{2 \pi} Q^2}  \int_0 ^{\infty} Dy  P(\xi y) \nonumber\\
\hspace*{-20mm}
&& \times \{
2 \sqrt{1-\xi ^2} \xi ^2 y +\sqrt{2 \pi} [\xi^2+(1- \xi^2)\xi ^2 y^2]
 e^{ \frac{1-\xi ^2}{2}y^2}[1-2H( \sqrt{1-\xi ^2}y)]\} \label{eq:bcrime}.
\end{eqnarray}
In the expression (\ref{eq:bcrime}), 
the first term in the parenthesis is ${\cal O}(\xi^{3+ \delta}/Q^2)$ and
the second term is evaluated as
$\frac{1}{ Q^2}  \int_0 ^{\infty} Dy  P( y)y^2$.
Thus, $A' - B' \simeq \frac{1}{ Q^2}  \int_0 ^{\infty} Dy (1-P(y))y^2$.
Then,
\begin{eqnarray}
\hat{q} & \simeq & \frac{\alpha}{ (\Delta q) ^2 }
\int_0 ^{\infty} Dy  y^2 [ 1 - P(y) ] \equiv
\frac{\alpha}{ (\Delta q) ^2 } g_3.
\end{eqnarray}
If $P(u)=1$, that is in the deterministic case, this gives 0.
Thus, we discuss this case later.
Similarly, from equation $(\ref{eq:s4})$ for $\hat{R}$
we obtain
\begin{eqnarray}
\hspace*{-10mm}
\hat{R} & \simeq & 
 \frac{\alpha}{\xi Q^2}
\{ \frac{2 \xi^4}{\sqrt{2\pi}}
 \int _0 ^{\infty} Dy P(\xi y)y 
 - \xi \sqrt{1-\xi^2}\int _0 ^{\infty} Dy P(y)
(1-y^2)[ 1 - 2H(\frac{ \sqrt{1-\xi^2}}{\xi} y)]\} \nonumber\\
\hspace*{-10mm}
&  \simeq &  \frac{\alpha}{\Delta q} \int _0 ^{\infty} Dy P(y)(y^2-1)
\equiv  \frac{\alpha}{\Delta q} g_4.
\end{eqnarray}
If $P(y)$ is not constant for $y>0$, the integration is positive.
If $P(y)$ is constant for $y>0$ , which can happen when $\delta=0$,
the integration is 0.  The latter case, we can perform the exact
calculation and obtain
\[
 \int _0 ^{\infty} Dy P(\xi y)\tilde{F}(y)
  = \frac{k \xi^4}{\pi}(2-\xi ^2) \simeq \frac{2k \xi^4}{\pi},
\]
where $k=P(y)$ for $y>0$.
Therefore,
\begin{equation}
\hat{R} \simeq \frac{\alpha}{\Delta q} \frac{2k \xi^3}{\pi}.
\end{equation}
Finally, let us consider the deterministic case.  In this case,
$\delta = 0$ and $k=1$ and $q=R$ and $ \hat{q}=\hat{R}$ hold.
 Then,
we obtain $\frac{\sqrt{1-\xi^2}}{\xi}=\frac{1}{Q}$
and $E(u/Q)=2H(u/Q)$.  Thus, equation $(\ref{eq:s3})$  becomes
\begin{equation}
 \hat{q} = \frac{2 \alpha}{1-q}\int Du 
\frac{h(u/Q)^2}{H(u/Q)} \simeq \frac{\alpha}{\sqrt{\Delta q}}
\frac{2}{\sqrt 2 \pi}\int Du \frac{h(u)}{H(u)}
\equiv \frac{\alpha}{\sqrt{\Delta q}} g_{3,D}.
\end{equation}
As for $\hat{R}$, since $v=-u$,
from equation $(\ref{eq:s3})$ we obtain exactly 
\begin{equation}
\hat{R}= \frac{2 \alpha}{1-q}\int Du 
\frac{h(u/Q)^2}{H(u/Q)}=\hat{q}.
\end{equation}

\section{ Asymptotic form of $I$ for $\tau \gg 1$ and $\hat{R} \gg 1$} 
 $I$ is expressed as
\begin{eqnarray}
I & = & \int Dt \ln [ 2{\rm cosh}( \sqrt{\hat{q}}t+\hat{R}) ]
 \simeq  \hat{R} + \frac{h(\tau)}{\tau \mu}+ I_1,\\
I_1 & = &I_1 ^- +I_1 ^+,\nonumber\\
 I_1 ^{\pm} & =&  \int _{\pm \tau} ^{\infty} Dt
\ln [ 1+ e^{-2 \sqrt{\hat{q}}(t \mp \tau)} ]
=\sqrt{2\pi} h(\tau)\int_0 ^{\infty}  Dx e^{ \mp x \tau}
\ln(1+ e^{-2\sqrt{\hat{q}}x}),\nonumber
\end{eqnarray}
where $\tau=\hat{R}/ \sqrt{\hat{q}}$ and 
$\mu=\frac{\hat{R}}{ 2 \hat{q}}$. 
The following relations are proved mathematically exactly,
\begin{eqnarray}
I_1 ^{\pm} &=& \sum_{n=1}^{\infty}\frac{(-1)^{n-1}}{n}
 e^{ 2 \hat{q} n^2 \pm 2 \hat{R} n } H
 [ \tau(\frac{n}{\mu} \pm 1) ]. \label{eq:b55}
\end{eqnarray}
For $0<\mu < 1$, in equations $(\ref{eq:b55})$,
$H(\tau ( \frac{n}{\mu}-1 ))$
and $H(\tau ( \frac{n}{\mu}+1 ))$
are approximated by
$\frac{h(\tau ( \frac{n}{\mu}-1 ))}
{\tau ( \frac{n}{\mu}-1 )}$
 and $\frac{h(\tau ( \frac{n}{\mu}+1 ))}
{\tau ( \frac{n}{\mu}+1 )}$, respectively.
Then,  $I_1$ is estimated as
\[
I_1  \simeq  \sum_{n=1}^{\infty}\frac{(-1)^{n-1}}{n}
h(\tau) [ \frac{1}{ \tau ( \frac{n}{\mu}-1 )} +  
\frac{1}{ \tau ( \frac{n}{\mu}+1)} ] = \frac{2 \mu }{ \tau} h(\tau) c(\mu),
\]
where $c(\mu) \equiv 
\sum_{n=1}^{\infty} \frac{(-1)^{n-1}}{n^2}
\frac{1}{1-(\mu /n)^2}$.
When $\mu$ is not an integer, 
\[
c(\mu)=\frac{\pi}{2\mu \sin(\mu \pi)}-\frac{1}{2\mu ^2}.
\]
Thus,
\begin{eqnarray*}
 I &\simeq& \hat{R} + \frac{h(\tau)}{ \tau \mu}
[ 1 + 2 \mu ^2 c(\mu)] = \hat{R} +
\frac{\psi(\mu) h(\tau) }{\tau \mu} \; \mbox{ for $0< \mu < 1$},\\
&& \psi(\mu) \equiv 1 + 2 \mu ^2 c(\mu) 
= \frac{\pi \mu}{\sin(\pi \mu)}.
\end{eqnarray*}
When $\mu =0$, $c(0)=\pi ^2/12$ and $\psi(0)=1$.  Then,
\[
I \simeq \hat{R} + \frac{ h(\tau) }{\tau \mu} 
\mbox{ for $\mu \simeq 0$}.
\]
 For $ \mu \ge 1$,  $I_1 ^-$  is expressed as
\begin{eqnarray}
\hspace*{-20mm}
I_1^- & \simeq & 
 \sum_{n=1}^{n_0}\frac{(-1)^{n-1}}{n}
 e^{ 2 \hat{q} n^2 - 2 \hat{R} n }
\{1- H [- \tau(\frac{n}{\mu} -1) ]\} 
 +  \frac{h(\tau)}{\tau} \sum_{ n =n_0 +1 }^{\infty}\frac{(-1)^{n-1}}{n}
 \frac{1}{ \frac{n}{\mu}-1} \label{eq:b12}, 
\end{eqnarray}
where $n_0 =[\mu ]$, i.e., $n_0$ is the largest integer which
does not exceed the value of $\mu$.
Let us compare terms in equation $(\ref{eq:b12})$.
Let us assume $1 \le n_1 < n_2 \le n_0$.
Then,
\[
 e^{ 2 \hat{q} n_1 ^2 - 2 \hat{R} n_1 }/
 e^{ 2 \hat{q} n_2 ^2 - 2 \hat{R} n_2 }=
 e^{ 4 \hat{q}( n_2- n_1)(\mu - (n_2- n_1)/2)}
> e^{ 4 \hat{q}( n_2- n_1)(n_0- n_1)}.
\]
Thus, we obtain
\[
 e^{ 2 \hat{q} n_1 ^2 - 2 \hat{R} n_1 } \gg
 e^{ 2 \hat{q} n_2 ^2 - 2 \hat{R} n_2 } \hbox{ for $  \hat{q} \gg 1$}.
\]
$  \hat{q} \gg 1$ is satisfied when $\hat{R} \gg 1$ 
as long as $\mu =\frac{\hat{R}}{2 \hat{q}}$
is bounded from the above.
Further, since
$ e^{ - \tau ^2/2 }/ e^{2\hat{ q}n^2 -2\hat{R}n}
 = e^{-2\hat{ q}(n-\mu)^2},
$ each term in the first summation in equation $(\ref{eq:b12})$ 
is lower order than $h(\tau)/\tau$ for $\tau \gg 1$.
Thus, $I_1 ^+$ and the second term in $I_1 ^-$
 are higher order than terms in the first term in $I_1 ^-$.
 Then, for $1 < \mu <2$,
\begin{eqnarray*}
\hspace*{-10mm}
I_1^- & \simeq &  e^{-2(\hat{R} - \hat{q})} 
 + \frac{2 h(\tau)\mu}{\tau}c ^- (\mu),\;\;
I_1^+  \simeq   \frac{2 h(\tau)\mu}{\tau}c ^+ (\mu),\\
\hspace*{-10mm}
&& c ^{\pm} (\mu) \equiv \frac{1}{2 \mu}
\sum_{n=1}^{\infty} \frac{(-1)^{n-1}}{n}
 \frac{1}{ \frac{n}{\mu} \pm 1 }.
\end{eqnarray*}
Thus,
\begin{eqnarray*}
 I_1 & \simeq & e^{-2(\hat{R} - \hat{q})}
+\frac{2 h(\tau)\mu}{\tau}c (\mu),
\end{eqnarray*}
where $ c (\mu) = c ^- (\mu) + c ^+  (\mu)$.
Therefore,
\[
I  \simeq  \hat{R}+e^{-2(\hat{R} - \hat{q})}
+\frac{h(\tau) \psi(\mu)}{\tau \mu} \mbox{   for $1< \mu <2$}.
\]
For $\mu=1$, we obtain
\[
 I_1  \simeq  e^{-2(\hat{R} - \hat{q})}/2
+\frac{2 h(\tau)\mu}{\tau}c_2 (\mu),
\]
where $c_2 (\mu) $ is defined as
\[
 c_2 (\mu) \equiv \frac{1}{2 \mu}
\sum_{n=2}^{\infty} \frac{(-1)^{n-1}}{n} \frac{1}{ \frac{n}{\mu}-1 }
+ c ^+ (\mu) = c(\mu) -\frac{1}{2(1-\mu)}.
\]
Thus,
\[
I  \simeq  \hat{R}+e^{-2(\hat{R} - \hat{q})}/2  \mbox{  for  $\mu = 1$}.
\]
We use the facts that $c_2(\mu)$ is analytic at $\mu=1$ and $c_2(1)=0$.\\
For $\mu \ge 2$, 
\begin{eqnarray*}
I_1^- & \simeq &  e^{-2(\hat{R} - \hat{q})} -\frac{1}{2}
 e^{-4(\hat{R} - 2\hat{q})} H[ \tau(\frac{2}{\mu} -1)],
\end{eqnarray*}
then,
\begin{eqnarray*}
 I_1 & \simeq & I_1^- \simeq e ^{-2(\hat{R} - \hat{q})} 
-\frac{1}{2} b_{\mu} e ^{-4(\hat{R} - 2\hat{q})},\\
&& b_{\mu}= 1 \hbox{ for $\mu>2$},\;\;
 b_2 = 1/2.
\end{eqnarray*}
Thus,
\[
I \simeq \hat{R}+ e ^{-2(\hat{R} - \hat{q})}
-\frac{1}{2} b_{\mu} e ^{-4(\hat{R} - 2\hat{q})} \mbox{ for $\mu \ge 2 $}.
\]

In summary, up tp the second order term in $I$, we obtain
\begin{eqnarray}
I & \simeq & \hat{R} + \frac{ h(\tau)\psi(\mu) }{\tau \mu}
\mbox{ for $0 \llsim \mu < 1$}, \label{eq:b5}\\
I & \simeq & \hat{R} + a_{\mu} e ^{-2(\hat{R} - \hat{q})}
\mbox{ for $1 \le \mu$},\\
&& \psi(\mu) \equiv 1 + 2 \mu ^2 c(\mu)
= \frac{\pi \mu}{\sin(\pi \mu)},\;\;
 a_1 = 1/2, \;\; a_{\mu} =1 \mbox{ for $\mu > 1$}.
\end{eqnarray}

\section{ $S_{PL}=0 $ and $f_{PL}=\alpha \epsilon_{min}$} 
As is shown in equation (3.84), when $q \rightarrow 1$ and
$R \rightarrow 1$ for $0<\beta <\infty$ and
for any $\delta$ and any $\chi$, $S_{RS}$ is expressed as
\begin{eqnarray}
S_{RS} &=&-\frac{\hat{q} \Delta q}{2} -(1 - \Delta R) \hat{\
R}
+ \alpha \sqrt{\Delta q}\; r(\chi,\beta) + I.
\end{eqnarray}
We consider the case of $0<\beta<\infty$ and $ 0 \le \delta <1/2$.\\
For the PL, $R=q=1, \chi=1$ and $\xi=Q=\sqrt{\Delta q}$.
Then,
\begin{eqnarray}
\hat{q} & = & \frac{\alpha}{\sqrt{\Delta q}}
g_{1, \delta}(\chi = 1, \beta),\label{eq:c2}\\
\hat{R} & = & \alpha (\Delta q) ^{ (\delta-1)/2}
g_{2, \delta}(\chi = 1, \beta).\label{eq:c3}
\end{eqnarray}
For $0<\beta<\infty$ and $\delta \ge 0$,
$g_{1, \delta}(1, \beta)$ and $g_{2, \delta}(1, \beta)$
 are finite. Then,
 $\mu=\frac{\hat{R}}{2\hat{q}}
=\frac{g_{2, \delta}}{2g_{1, \delta}}(\Delta q)^{\delta/2}$
becomes
\begin{eqnarray*}
\mu \simeq 0 \mbox{ for } \delta>0,\\
\mu =\frac{g_{2, \delta}}{2g_{1, \delta}}
= \mbox{finite  for } \delta=0.
\end{eqnarray*}
Let us estimate $I$.
In the case of $\delta>0$, since $\mu =0$ we obtain from $(\ref{eq:b5})$
\[
I \simeq \hat{R}+ \frac{ h(\tau)}{ \tau \mu} =
\hat{R}+ \frac{2 \hat{R} }{ \tau ^3} h(\tau).
\]
  Then,
\[
S=- \frac{1}{2}g_{1, \delta} \alpha \sqrt{\Delta q}
+ g_{2, \delta} \alpha (\Delta q)^{\frac{1+ \delta}{2}}
+ \frac{2 \hat{R} }{ \tau ^3} h(\tau)
+ \alpha r  \sqrt{\Delta q}.
\]
From the equations $(\ref{eq:c2})$ and $(\ref{eq:c3})$, we obtain
\[
 \hat{q}^{\delta -1} \hat{R} = (\alpha g_{1, \delta})^{\delta -1}
\alpha g_{2, \delta} \equiv C.
\]
Then, 
\[
  \hat{R} = C \hat{q}^{1 - \delta},\;\;
  \frac{\hat{R}}{\tau ^3} = C ^{- \frac{1}{1 - 2 \delta}}
\tau ^{ \frac{4 \delta -1}{1 - 2 \delta}}.
\]
Since $\Delta q=0, \tau =\infty$ and $g_{1, \delta}, 
g_{2, \delta} $, $r$ and $C$ are finite, we obtain $S=0.$
For the case of $\delta=0$, we have to determine 
the value of $\mu$.  As is discussed in \S3,
when $\chi$ is finite, $\mu = \frac{1}{1+ \chi ^2}$
from equation  $(\ref{eq:rsmua})$.  
In our case, $\chi=1$ and then $\mu = 1/2$.
Thus,
\[
I \simeq \hat{R}+ \frac{\pi h(\tau) }{\tau}.
\]
By a similar argument to the case of $\delta>0$, we obtain $S=0.$
Now, let us estimate $<e_t>= - \alpha e^{-\beta}J$.
$J$ is estimated as
\begin{eqnarray*}
J
%& =& \int Dy2{\cal P}(y) \int Du
%\frac{H(Y)-1}{\tilde{H}(Y)} \\
&=& \int Du \frac{H(u/Q)-1}{\tilde{H}(u/Q)}
\int Dy[1-P(\xi y + \sqrt{1- \xi ^2} u)]\\
&=& \int Du \frac{H(u/Q)-1}{\tilde{H}(u/Q)}
[1-P(u)]= - e^{\beta} \int _0 ^{\infty}Du[1-P(u)]
= - e^{\beta}\epsilon_{min}.
\end{eqnarray*}
Thus, 
$<e_t>= - \alpha e^{-\beta}J= \alpha \epsilon_{min}.$
Then, we obtain
$f_{PL} =  <e_t>- TS_{PL}= \alpha \epsilon_{min}.$

\section{ Asymptotic form of $f_{RS}$ and $S_{RS}$ for $\beta \ll 1.$} 
In this appendix, we derive
the asymptotic forms of the free energy 
and the entropy for the RS solution for $\beta \ll 1.$\\
$f_{RS}$ is expressed as
\begin{eqnarray}
&&  -{\beta} f_{RS}(q,\hat{q},R,\hat{R},\beta) =
  -\frac{\hat{q}}{2} (1- q) - R \hat{R} + \alpha K + I, \\
&& \hspace{0.5cm} K \equiv \int Dy2{\cal P}(y)
 \int Du \ln \tilde{H}(Y), \;\;
 I \equiv \int Dt 
\ln [2 {\rm cosh}( \sqrt{\hat{q}}t+ \hat{R})], \nonumber 
\end{eqnarray}
where $Y= \frac{\sqrt{q-R^2}u-Ry}{\sqrt{1-q}}$.\\
  By defining $K_a$ and $K_b$ as follows,
\begin{eqnarray*}
K_a & \equiv& \int Dy2{\cal P}(y) \int Du H(-Y) =
%\nonumber \\
%& = & 
\epsilon_{min} + 2 \int _0 ^{\infty} Dy P(y) H(\frac{Ry}{\sqrt{1-R^2}})
=\epsilon_g, \\
K_b & \equiv & \int Dy2{\cal P}(y) \int Du H(Y) H(-Y) 
 =   Q \int \tilde{D}uH(u)H(-u),
\end{eqnarray*}
 $K$  and $ f_{RS}$ are expressed as
\begin{eqnarray*}
K & = & - \beta (K_a - \frac{\beta}{2}K_b) + {\cal O}(\beta ^3),\\
  -{\beta} f_{RS} & = &
  -\frac{\hat{q}}{2} (1- q) - R \hat{R} + I
  - \alpha \beta (K_a - \frac{\beta}{2}K_b) + {\cal O}(\beta ^3) \nonumber \\
& \simeq &  -\frac{\hat{q}}{2} (1- q) - R \hat{R} + I
  - \alpha \beta \epsilon_g + \frac{\alpha \beta^2}{2}K_b.
\end{eqnarray*}
The entropy $S_{RS}$ is expressed as
\begin{eqnarray}
S_{RS} & = & -\frac{\hat{q}}{2} (1- q) - R \hat{R} +  I
+\alpha K - \alpha \beta e^{-\beta} J, \\
&& \hspace{0.5cm} J = \int Dy2{\cal P}(y) \int Du
\frac{H(Y)-1}{\tilde{H}(Y)}\nonumber.
\end{eqnarray}
Then, defining $L$ as $L = K -  \beta e ^{-\beta} J$,
we get
\[
 L=- \frac{\beta ^2}{2}K_b +{\cal O}(\beta ^3).
\]
Thus, we obtain
\[
 S_{RS} =  -\frac{\hat{q}}{2} \Delta q -R \hat{R}+I
-\frac{\alpha \beta ^2}{2}K_b +{\cal O}(\beta ^3).
\]
For $\Delta q \ll 1$ and $\Delta R \ll 1$,
\begin{eqnarray*}
K_a &=&  \epsilon_g \simeq 
\epsilon_{min} + \frac{2s}{(1+\delta) \sqrt{2\pi}}
(2\Delta R)^{\frac{1+\delta}{2}},\;\;
K_b \simeq   \frac{\sqrt{\Delta q}}{\pi \sqrt{2} }.
\end{eqnarray*}
Then, $K$, $f_{RS}$ and $S_{RS}$ are expressed as
\begin{eqnarray}
\hspace{-20mm}
 K & = & -\beta [ \epsilon_{min} +
\frac{2s}{(1+\delta) \sqrt{2\pi}}
(2 \Delta R)^{\frac{1+\delta}{2}} 
- \frac{ \beta \sqrt{\Delta q}}{2 \pi \sqrt{2} } ], \\
\hspace{-20mm}
  -{\beta} f_{RS} & = &
  -\frac{\hat{q}}{2} (1- q) - R \hat{R} + I 
  - \alpha \beta [ \epsilon_{min} +
\frac{2s}{(1+\delta) \sqrt{2\pi}}
(2 \Delta R)^{\frac{1+\delta}{2}}
- \frac{ \beta \sqrt{\Delta q}}{2 \pi \sqrt{2} } ], \\
\hspace{-20mm}
 S_{RS}& =&  -\frac{\hat{q}}{2} \Delta q 
- \hat{R} R + I
- \alpha \beta ^2 \frac{\sqrt{\Delta q}}{2 \pi \sqrt{2} }.
\end{eqnarray}

\end{document}

\def\vec0{\mbox{\boldmath $w^0$}}
\def\vecx{\mbox{\boldmath $x$}}
\def\vecw{\mbox{\boldmath $w$}}
\def\veceta{\mbox{\boldmath $\eta$}}

{\bf Figure captions}

Fig.1 $\alpha$ dependence of several quantities in the
 RS solution for $\delta=1$ and $T=1$. \\
(a) $q$ and $R$(dotted curve) (b) $S_{RS}$ 
(c) $\Lambda_1$ and $\Lambda_3$(dotted curve)

Fig.2 $\alpha$ dependence of $q$ and $R$
(dotted curve) in
the RS solution for $\delta=0$. \\
%(a) $T=5$ (b) $T=1$ (c) $T=0.5$
(a) $T=5$ (b) $T=0.5$

Fig.3 $\alpha$ dependence of the entropy in the RS solution for $\delta=0$. 
Solid and dashed curves denote $S_{RS}^I$ and $S_{RS}^{II}$,
respectively.\\
%(a) $T=5$ (b) $T=1$ (c) $T=0.5$
(a) $T=5$ (b) $T=0.5$

Fig.4  $\alpha$ dependence of $\Lambda_1$ and $\Lambda_3$
(dotted curve)
 in the RS solution for $\delta=0$.
$\Lambda_1 ^I$ and $\Lambda_3^I$ 
correspond to curves starting from -1 for small $\alpha$.\\
%(a) $T=5$ (b) $T=1$ (c) $T=0.5$
(a) $T=5$ (b) $T=0.5$

Fig.5 $T$ dependence of $q(\alpha)$ and $R(\alpha)$(dotted curve)
in the RS solution for  $T=5$.\\
%(a) $\delta=0.3$  (b) $\delta=0.5$ (c) $\delta=1.5$
(a) $\delta=0.3$  (b) $\delta=1.5$

Fig.6  $\alpha$ dependence of $T_c $ for $\delta=1$.
% (a) $q (RS), q_0 (1RSB)$ (b) $R$ (c) $\Delta \epsilon_g $ (d) 

Fig.7   $\ln \alpha $ v.s. $\ln \Delta \epsilon_g $
 A line segment with estimated gradient is depicted togehter.\\

Fig.8  $\alpha$ dependence of  $T_c$ for $\delta=0$. 
 Solid curve: 1RSB(I), dashed curve: 1RSB(II).\\
%(a) $q(RS), q_0(1RSB)$ (b) $R$ (c) $\Delta \epsilon_g$ (d)

Fig.9 $\alpha$ dependence of Free energy. $f^I_{RSB}$(solid curve)
 and $f^{II}_{RSB}$ ( dashed curve)

Fig.10  $\alpha$ dependence of $R$ and $\delta R$ 
in the minimum-error algorithm for $\delta=1$. 
Dotted curve: $N=10$, Dashed curve: $N=15$, Solid curve: $N=17$.\\
(a) $R$ (b) $\delta R$

Fig.11  $\alpha$ dependence of several quantites
in the minimum-error algorithm  for $\delta=1$. 
+: numerical results for $N=17$, bars indicate standard deviations, 
Dashed curve: RS solution, Solid curve: 1RSB solution \\
(a) $R$ (b) $q$(RS) and $q_0$(1RSB) (c) $\Delta \epsilon_g$

Fig.12  Asymptotic behavior of $R$ in the minimum-error algorithm  
for $\delta=1$.
+: numerical results for $N=15$, bars indicate standard deviations, 
Dashed curve: RS solution, Solid curve: 1RSB solution\\
(a) $0<\alpha<15$  (b)$15<\alpha<50$ (c)$50<\alpha<100$

Fig.13  $\alpha$ dependence of  $R$ and $\Delta \epsilon _g $
in the Gibbs algorithm for $\delta=1$. 
+: numerical results for $N=12$ with $T=1$,
% bars indicate standard deviations, 
dashed curve: RS solution with $T=0$, 
dotted curve: RS solution with $T=1$, 
solid curve: 1RSB solution\\
(a) $R$ (b) $\Delta \epsilon _g $

Fig.14  $\alpha$ dependence of $q$ and $q_0$
in the Gibbs algorithm for $\delta=1$. 
+: numerical results with standard deviations for $N=12$,
dashed curve $q$: RS solution with $T=0$,
dotted curve $q$: RS solution with finite temperature,
solid curve $q_0$ :1RSB solution\\
%(a) $T=0.15$, (b) $T=0.5$ (c) $T=1.0$ (d) $T=5.0$
(a) $T=0.15$, (b) $T=0.5$ (c)  $T=5.0$

Fig.15  $T$ dependence of $P(q)$ in the Gibbs algorithm for $\delta=1$.
Histgram: numerical results for $N=12$ and $p=60$, 
solid line: 1RSB solution, dotted line: RS solution.\\
%(a) $T=0.15$ (b) $T=0.5$ (c) $T=1.0$ (d) $T=5.0$
(a) $T=0.15$ (b) $T=0.5$ (c) $T=5.0$

Fig.16  $\alpha$ dependence of several quantites
in the minimum-error algorithm  for $\delta=0$.
+: numerical results for $N=17$, bars indicate standard deviations, 
Dashed curve: RS($T=0$), solid curve: 1RSB(I), dashed curve: 1RSB(II).\\
(a) $R$ (b) $q$(RS) and $q_0$(1RSB) (c) $\Delta \epsilon_g$

Fig.17  Asymptotic behavior of $R$ in the minimum-error algorithm  
for $\delta=0$.
+: numerical results for $N=15$, bars indicate standard deviations, 
Dashed curve: RS solution, Solid curve: 1RSB solution

Fig.18  $\alpha$ dependence of $R$ and $\Delta \epsilon_g$
in the Gibbs algorithm for $\delta=1$ and $T=1.0$.
+: numerical results for $N=12$.
% bars indicate standard deviations, 
dashed curve: RS solution with $T=0$, 
dotted curve: RS solution for $T=1.0$.
solid curve: 1RSB solution\\
(a) $R$ (b) $\Delta \epsilon_g$

Fig.19 $\alpha$ dependence of $q$ and $q_0$
in the Gibbs algorithm for $\delta=0$. 
+: numerical results with standard deviations for $N=12$,
dashed curve: RS solution with $T=0$, 
dotted curve: RS solution with finite temperature,
solid curve: 1RSB solution\\
%(a) $T=0.15$ (b) $T=0.5$ (c) $T=1.0$ (d) $T=5.0$
(a) $T=0.15$ (b) $T=0.5$ (c) $T=5.0$

Fig.20  $T$ dependence of $P(q)$ 
in the Gibbs algorithm for $\delta=0$.
$T_c \simeq 0.7$. Histgram : numerical results for $N=12$ and $p=60$, 
solid line: 1RSB solution, dotted line: RS solution.\\
%(a) $T=0.15$ (b) $T=0.5$ (c) $T=1.0$ (d) $T=5.0$
(a) $T=0.15$ (b) $T=0.5$ (c) $T=5.0$

% Fig. 1

\begin{figure}
  \caption{Fig.1  $\alpha$ dependence of several quantities in the
 RS solution for $\delta=1$ and $T=1$. }
  \epsfxsize = 4cm
\centerline{\epsfbox{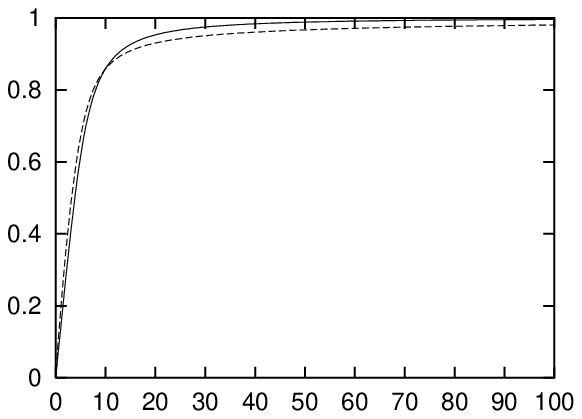}}
  \caption{Fig.1(a) $q$ and $R$(dotted curve)}
  \epsfxsize = 4cm
\centerline{\epsfbox{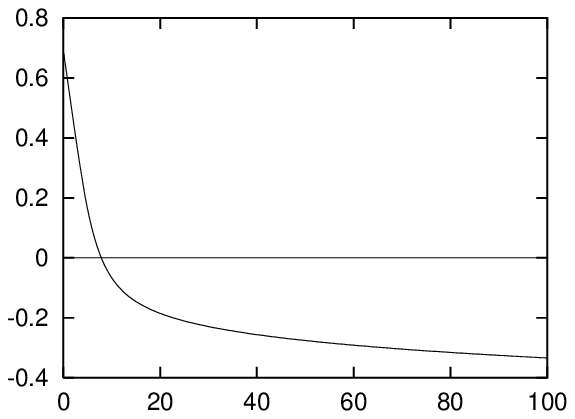}}
 \caption{Fig.1 (b) $S_{RS}$ }

  \epsfxsize = 4cm
\centerline{\epsfbox{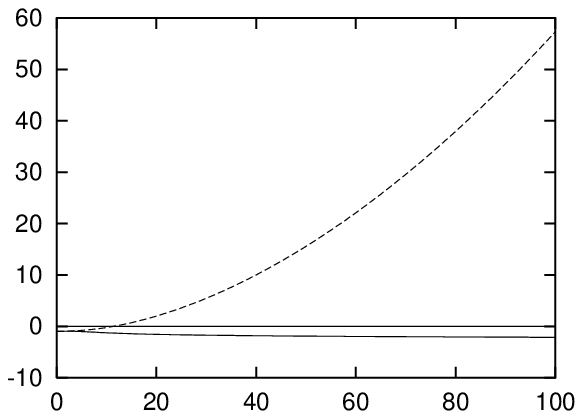}}
 \caption{Fig.1 (c) $\Lambda_1$ and $\Lambda_3$(dotted curve)}

\end{figure}

% Fig. 2

\begin{figure}
\caption{Fig.2 $\alpha$ dependence of $q$ and $R$
(dotted curve) in
the RS solution for $\delta=0$.}
  \epsfxsize = 4cm
   \centerline{\epsfbox{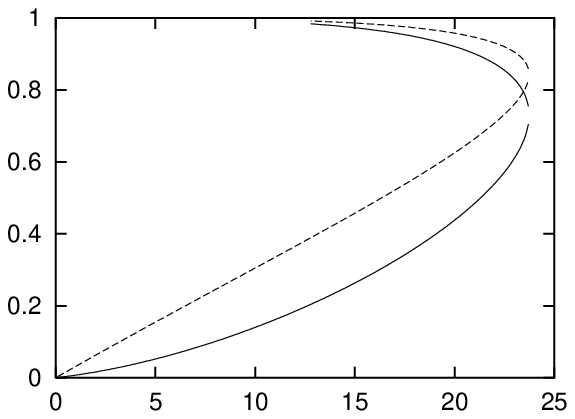}}
  \caption{Fig.2(a) $T=5$  }

%  \epsfxsize = 4cm
%   \centerline{\epsfbox{Uezu02b.eps}}
% \caption{Fig.2(b)$T=1$ }

  \epsfxsize = 4cm
   \centerline{\epsfbox{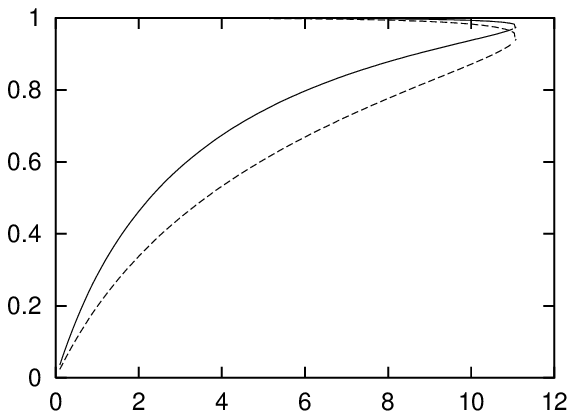}}
 \caption{Fig.2(b) $T=0.5$}
\end{figure}

% Fig. 3

\begin{figure}
\caption{ Fig.3  $\alpha$ dependence of the entropy in the RS solution for $\delta=0$. 
Solid and dashed curves denote $S_{RS}^I$ and $S_{RS}^{II}$,
respectively.
}
  \epsfxsize = 4cm
   \centerline{\epsfbox{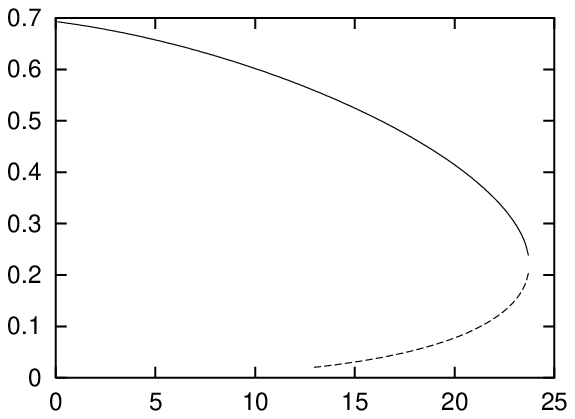}}
 \caption{Fig.3(a) $T=5$ }

%  \epsfxsize = 4cm
%   \centerline{\epsfbox{Uezu03b.eps}}
% \caption{Fig.3(b) $T=1$}

  \epsfxsize = 4cm
   \centerline{\epsfbox{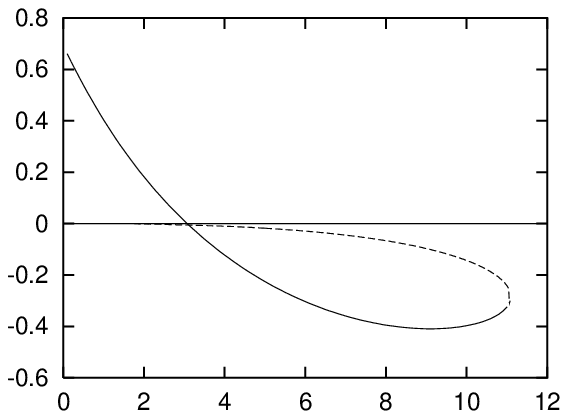}}
 \caption{Fig.3(b) $T=0.5$}

\end{figure}

% Fig. 4

\begin{figure}
\caption{ Fig.4 $\alpha$ dependence of $\Lambda_1$ and $\Lambda_3$
(dotted curve)
 in the RS solution for $\delta=0$. 
$\Lambda_1 ^I$ and $\Lambda_3^I$
correspond to curves starting from -1 for small $\alpha$.
}
  \epsfxsize = 4cm
   \centerline{\epsfbox{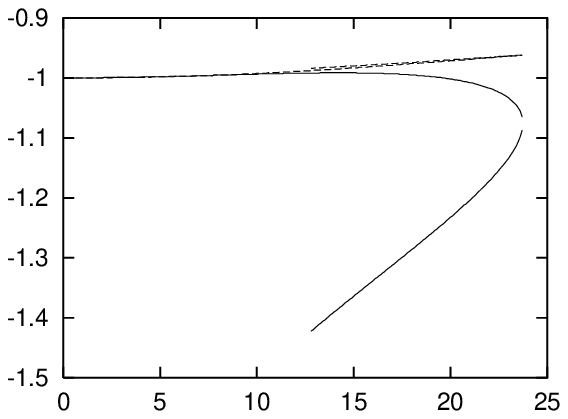}}
 \caption{Fig.4(a) $T=5$}

%  \epsfxsize = 4cm
%   \centerline{\epsfbox{Uezu04b.eps}}
% \caption{Fig.4(b) $T=1$}

  \epsfxsize = 4cm
   \centerline{\epsfbox{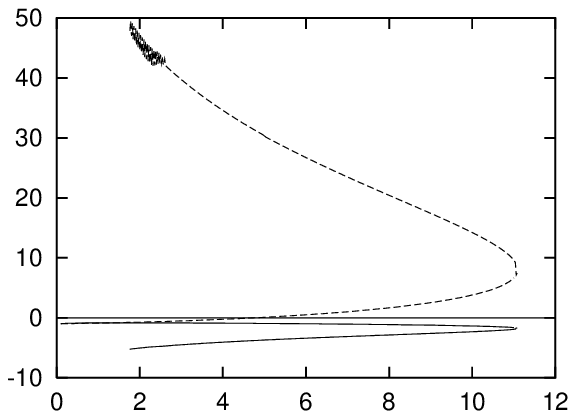}}
 \caption{Fig.4(b) $T=0.5$}

\end{figure}

%Fig.5

\begin{figure}
\caption{Fig.5 $\alpha$ dependence of $q$ and $R$(dotted curve)
in the RS solution for  $T=5$.
}
  \epsfxsize = 4cm
   \centerline{\epsfbox{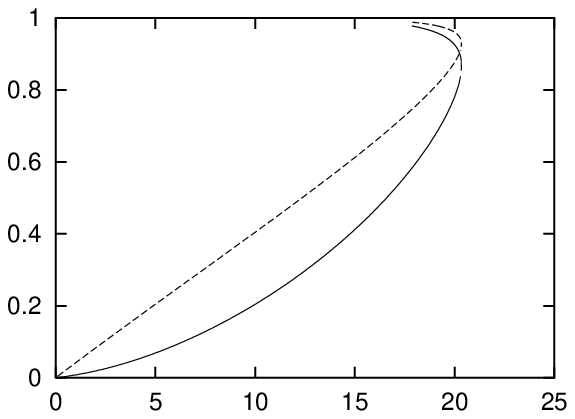}}
 \caption{Fig.5(a)$\delta=0.3$}

%  \epsfxsize = 4cm
%   \centerline{\epsfbox{Uezu05b.eps}}
% \caption{Fig.5(b) $\delta=0.5$}

  \epsfxsize = 4cm
   \centerline{\epsfbox{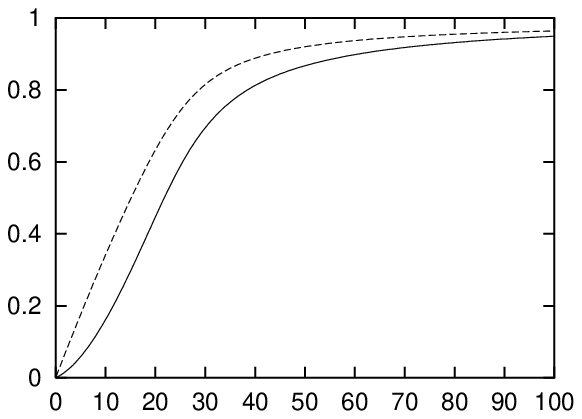}}
 \caption{Fig.5(b) $\delta=1.5$}

\end{figure}

% Fig. 6

\begin{figure}
%\caption{ Fig.6 $\alpha$ dependence of several quantites in the RS and the
%1RSB solutions for $\delta=1$. Solid curve: 1RSB, dashed curve:  RS($T=0$).
%}

%  \epsfxsize = 4cm
%   \centerline{\epsfbox{Uezu06a.eps}}
% \caption{Fig.6 (a) $q (RS), q_0 (1RSB)$ }

%  \epsfxsize = 4cm
%   \centerline{\epsfbox{Uezu06b.eps}}
% \caption{Fig.6 (b) $R$  }

%  \epsfxsize = 4cm
%   \centerline{\epsfbox{Uezu06c.eps}}   
% \caption{Fig.6 (c)$\Delta \epsilon_g $}

  \epsfxsize = 4cm
  \centerline{\epsfbox{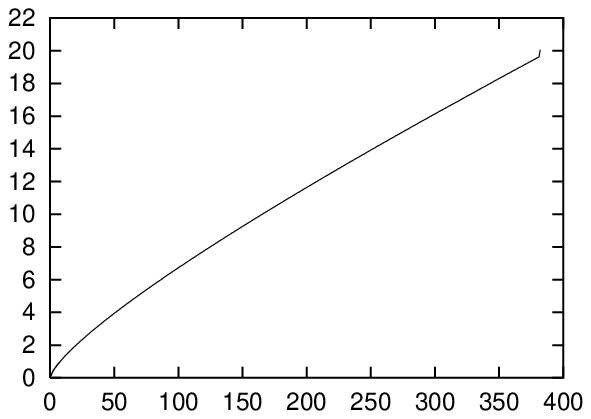}}
%  \caption{Fig.6  $T_c $ }
   \caption{ Fig.6 $\alpha$ dependence of $T_c$ for $\delta=1$.}
\end{figure}

% Fig. 7

\begin{figure}

  \epsfxsize = 4cm
      \centerline{\epsfbox{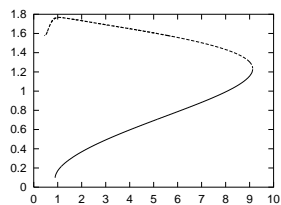}}
 \caption{Fig.7  $\ln \alpha $ v.s. $\ln \Delta \epsilon_g $
 A line segment with estimated gradient is depicted togehter.}

\end{figure}

% Fig. 8

\begin{figure}
%\caption{ Fig.8 $\alpha$ dependence of several quantites in the RS and the
%1RSB solutions for $\delta=0$.  Solid curve: 1RSB(I), dashed curve: 1RSB(II),
% dotted curve: RS($T=0$).
%}
%  \epsfxsize = 4cm
%      \centerline{\epsfbox{Uezu08a.eps}}
% \caption{Fig.8 (a)$q(RS), q_0(1RSB)$ }

%  \epsfxsize = 4cm
%      \centerline{\epsfbox{Uezu08b.eps}}
% \caption{Fig.8 (b) $R$ }

%  \epsfxsize = 4cm
%      \centerline{\epsfbox{Uezu08c.eps}}
% \caption{Fig.8 (c) $\Delta \epsilon_g$ }

  \epsfxsize = 4cm
      \centerline{\epsfbox{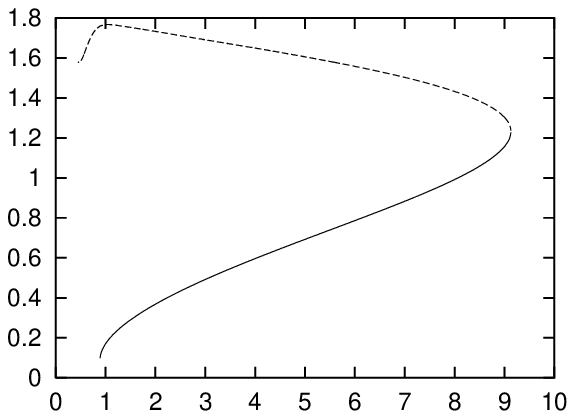}}
% \caption{Fig.8 (d)  $T_c$}
\caption{ Fig.8 $\alpha$ dependence of $T_c$ for $\delta=0$.
Solid curve: 1RSB(I), dashed curve: 1RSB(II).}
\end{figure}

\newpage
% Fig. 9

\begin{figure}

  \epsfxsize = 4cm
      \centerline{\epsfbox{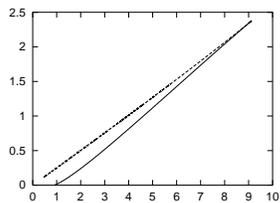}}
 \caption{Fig.9 $\alpha$ dependence of Free energy. $f^I_{RSB}$(solid curve)
 and $f^{II}_{RSB}$ ( dashed curve). }
\end{figure}
%\newpage

% Fig. 10
\begin{figure}
\caption{ Fig.10 $\alpha$ dependence of $R$ and $\delta R$ 
in the minimum-error algorithm for $\delta=1$. 
Dotted curve: $N=10$, Dashed curve: $N=15$, Solid curve: $N=17$.
}
  \epsfxsize = 4cm
      \centerline{\epsfbox{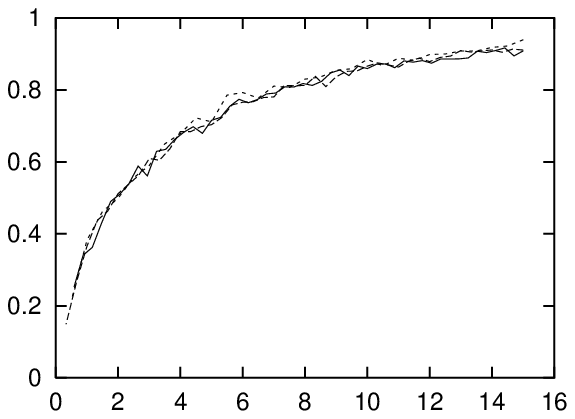}}
 \caption{Fig.10(a) $R$}

  \epsfxsize = 4cm
      \centerline{\epsfbox{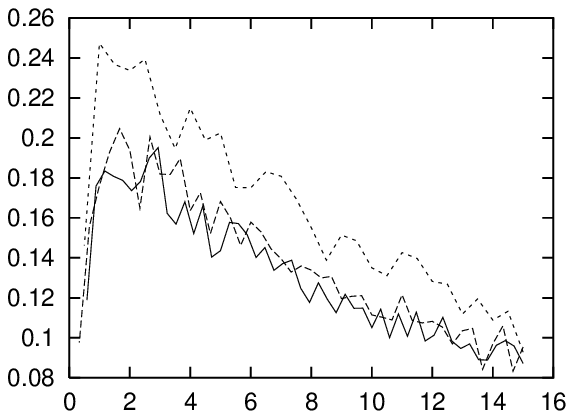}}
 \caption{Fig.10 (b) $\delta R$}

\end{figure}

\newpage

% Fig. 11

\begin{figure}
\caption{ Fig.11  $\alpha$ dependence of several quantites
in the minimum-error algorithm  for $\delta=1$. 
+: numerical results for $N=17$, bars indicate standard deviations, 
Dashed curve: RS solution, Solid curve: 1RSB solution
}
  \epsfxsize = 4cm
      \centerline{\epsfbox{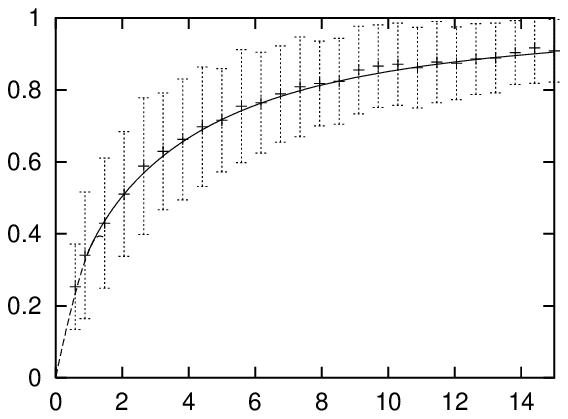}}
 \caption{Fig.11(a) $R$ }

  \epsfxsize = 4cm
      \centerline{\epsfbox{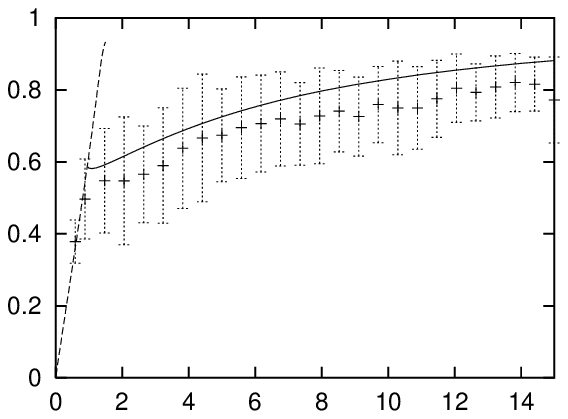}}
 \caption{Fig.11 (b) $q$(RS) and $q_0$(1RSB)}

  \epsfxsize = 4cm
      \centerline{\epsfbox{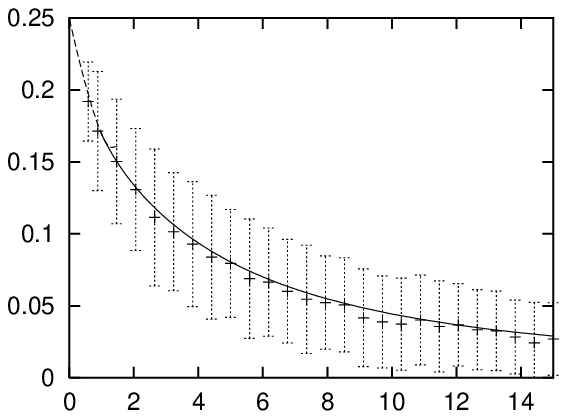}}
 \caption{Fig.11 (c) $\Delta \epsilon_g$}

\end{figure}

% Fig. 12

\begin{figure}
\caption{ Fig.12  Asymptotic behavior of $R$ in the minimum-error algorithm  
for $\delta=1$.
+: numerical results for $N=15$, bars indicate standard deviations, 
Dashed curve: RS solution, Solid curve: 1RSB solution
}
  \epsfxsize = 4cm
      \centerline{\epsfbox{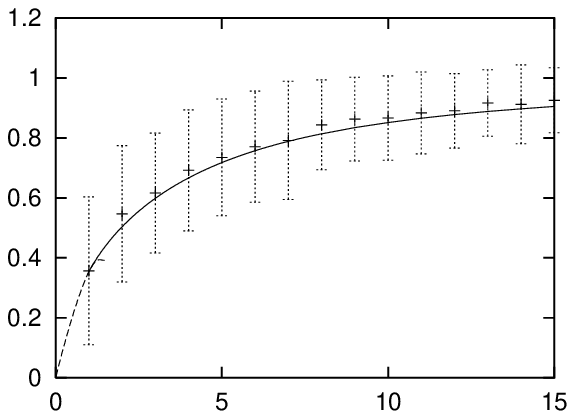}}
 \caption{Fig.12(a)  $0<\alpha<15$}

  \epsfxsize = 4cm
      \centerline{\epsfbox{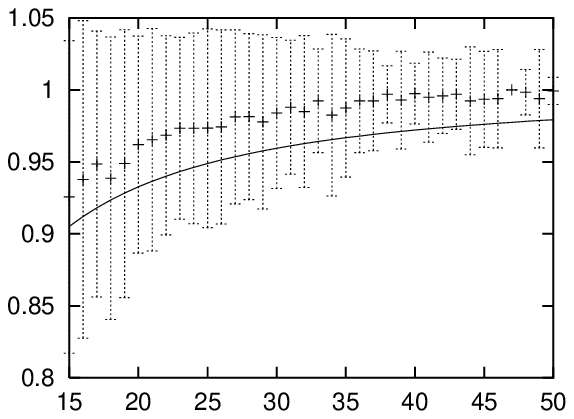}}
 \caption{Fig.12(b) $15<\alpha<50$ }

  \epsfxsize = 4cm
      \centerline{\epsfbox{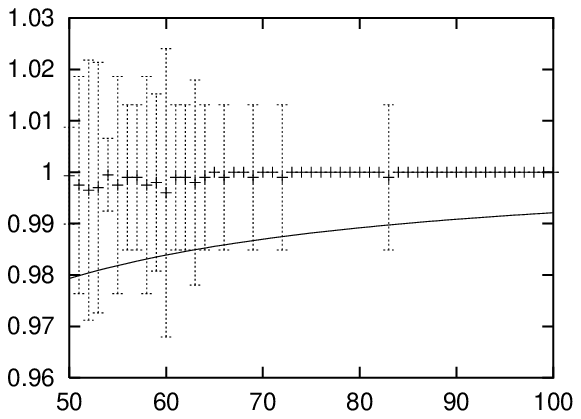}}
 \caption{Fig.12 (c) $50<\alpha<100$}

\end{figure}

% Fig. 13
\begin{figure}
\caption{Fig.13  $\alpha$ dependence of  $R$ and $\Delta \epsilon _g $
in the Gibbs algorithm for $\delta=1$. 
+: numerical results for $N=12$ with $T=1$,
% bars indicate standard deviations, 
dashed curve: RS solution with $T=0$, 
dotted curve: RS solution with $T=1$, 
solid curve: 1RSB solution
}
  \epsfxsize = 4cm
      \centerline{\epsfbox{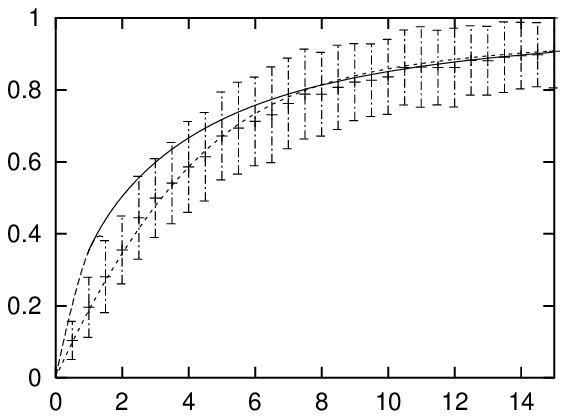}}
 \caption{Fig.13(a)  $R$}

  \epsfxsize = 4cm
      \centerline{\epsfbox{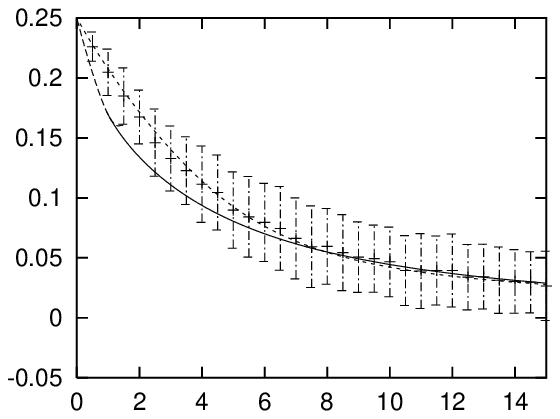}}
 \caption{Fig.13(b) $\Delta \epsilon _g $ }
\end{figure}

% Fig. 14
\begin{figure}
\caption{Fig.14  $\alpha$ dependence of $q$ and $q_0$
in the Gibbs algorithm for $\delta=1$. 
+: numerical results with standard deviations for $N=12$,
dashed curve $q$: RS solution with $T=0$, 
dotted curve $q$: RS solution with finite temperature,
solid curve $q_0$ :1RSB solution
}
  \epsfxsize = 4cm
      \centerline{\epsfbox{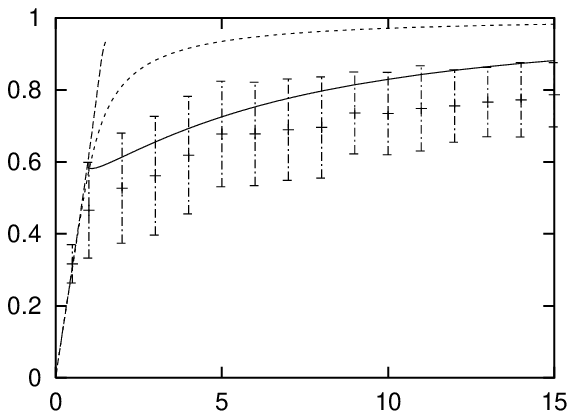}}
 \caption{Fig.14(a) $T=0.15$ }

  \epsfxsize = 4cm
      \centerline{\epsfbox{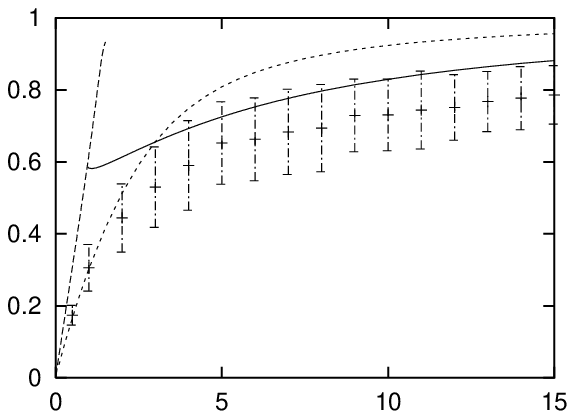}}
 \caption{Fig.14(b)  $T=0.5$}

%  \epsfxsize = 4cm
%      \centerline{\epsfbox{Uezu14c.eps}}
% \caption{Fig.14(c)  $T=1.0$}

  \epsfxsize = 4cm
      \centerline{\epsfbox{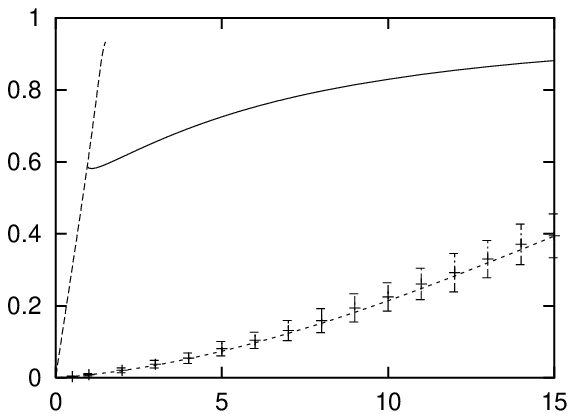}}
 \caption{Fig.14(c) $T=5.0$ }

\end{figure}

% Fig. 15

\begin{figure}
\caption{ Fig.15 $T$ dependence of $P(q)$ in the Gibbs algorithm for $\delta=1$.
Histgram: numerical results for $N=12$ and $p=60$, 
solid line: 1RSB solution, dotted line: RS solution.
}
  \epsfxsize = 4cm
      \centerline{\epsfbox{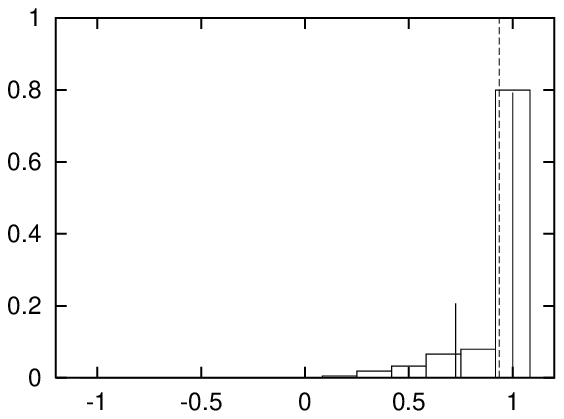}}
 \caption{Fig.15(a) $T=0.15$ }

  \epsfxsize = 4cm
      \centerline{\epsfbox{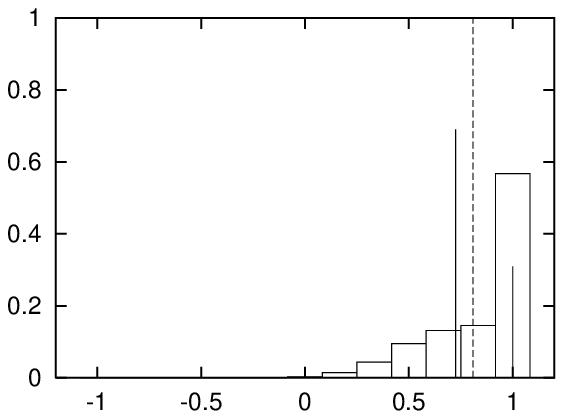}}
 \caption{Fig.15(b) $T=0.5$ }

%  \epsfxsize = 4cm
%      \centerline{\epsfbox{Uezu15c.eps}}
% \caption{Fig.15(c) $T=1.0$ }

 \epsfxsize = 4cm
      \centerline{\epsfbox{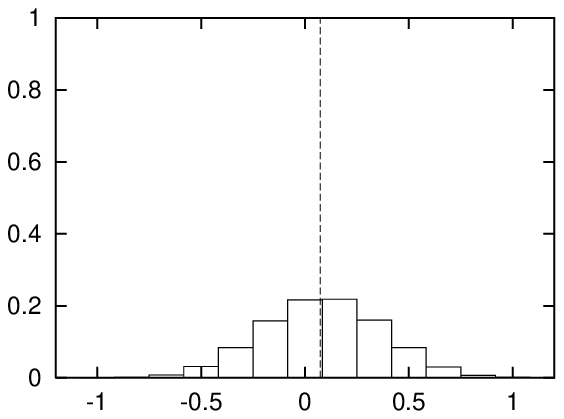}}
 \caption{Fig.15(c) $T=5.0$ }
\end{figure}